\documentclass[11pt]{article}
\usepackage{graphicx}  
\usepackage{amsmath} 
\usepackage{xfrac}
\usepackage{amssymb}   
\usepackage{bm} 
\usepackage{dcolumn}
\usepackage{ulem}
\usepackage{cite}
\usepackage{color}
\usepackage{mathtools}
\usepackage{mathrsfs}
\usepackage{amsfonts}
\usepackage{varioref}
\usepackage{textcomp}
\usepackage{geometry}
\usepackage{tcolorbox}
\usepackage{empheq}
\usepackage[active]{srcltx}
\usepackage{tikz}
\usetikzlibrary{er,positioning}
\usepackage{lmodern}
\usepackage{braket}
\usetikzlibrary{decorations.pathmorphing}
\tikzset{snake it/.style={decorate, decoration=snake}}
\usepackage{fancybox}

\RequirePackage[colorlinks,citecolor=blue,urlcolor=magenta,linkcolor=blue]{hyperref}
\allowdisplaybreaks

\def\be{\begin{equation}}
\def\bea{\begin{eqnarray}}
\def\ee{\end{equation}}
\def\eea{\end{eqnarray}}

\numberwithin{equation}{section}

\newcommand{\w}{\omega}
\newcommand{\Dp}{D_+}

\title{\bf  de Sitter Teukolsky waves}
\author{Harsh$^1$, Sk Jahanur Hoque$^{2,3,4}$,  Sitender Pratap Kashyap$^1$,   Amitabh Virmani$^1$}
\date{%
\small{  
\vskip 0.5cm    $^1$Chennai Mathematical Institute, H1, SIPCOT IT Park, Siruseri, \\ Kelambakkam 603103, Tamil Nadu, India} \\
\vskip 0.5cm   $^2$Birla Institute of Technology and Science, Pilani, Hyderabad Campus, \\ Jawaharnagar, Hyderabad 500 078, India \\
\vskip 0.5cm    $^3$Institute of Theoretical Physics, Faculty of Mathematics and Physics, \\ Charles University,  V Hole\v{s}ovi\v{c}k\'ach 2, 180 00 Prague 8, Czech Republic\\
\vskip 0.5cm    $^4$Universit\'{e} Libre de Bruxelles, International Solvay Institutes, \\ CP 231, B-1050 Brussels, Belgium\\
\vskip 0.5cm    {\small jahanur.hoque@hyderabad.bits-pilani.ac.in, \\ \{harsh22, sitender, avirmani\}@cmi.ac.in} \\
\vskip 0.5cm   \today \\
}

\begin{document}
\maketitle

\begin{abstract}
We present de Sitter Teukolsky waves---linearised quadrupolar gravitational waves in the transverse-traceless gauge in de Sitter spacetime. In the cosmological constant $\Lambda$ going to zero limit, our solutions match to Teukolsky solutions. For non-zero $\Lambda$, we compare our solutions to the wider literature, where different authors have constructed linearised gravitational perturbations in de Sitter spacetime with varied motivations. For de Sitter Teukolsky waves, we compute the energy flux across future timelike infinity $\mathcal{I}^{+}$ and show that it is manifestly positive.
\end{abstract}

\newpage

\tableofcontents

\newpage

\section{Introduction}

Although already in 1916 Einstein showed that general relativity admits gravitational waves in the weak field approximation, the exact nature of the gravitational waves in full non-linear theory of general relativity was a subject of a sustained controversy \cite{Kennefick:2007zz, Kennefick:1997kb}. The controversy was settled once and for all by the seminal works of Bondi, van der Burg, Metzner and Sachs \cite{Bondi:1962px, Sachs:1962wk}, who rigorously established the physical nature of gravitational waves.   
The Bondi-Sachs formalism (as it is now commonly known) describes  asymptotically flat spacetimes near future null infinity and captures all details of the asymptotic field generated by isolated self-gravitating sources.

Over the years, the Bondi-Sachs formalism has been explored from many different perspectives. For example, for a better understanding of the gravitational wave  generation, one needs the connection between the asymptotic structure of the fields and explicit multipole moments of the localised matter sources. With this motivation, the relation between the Bondi expansion and the multipolar expansion of the gravitational field is studied by many authors; see e.g., ref.~\cite{Blanchet:2020ngx} and references therein.

Another reason for wide interest in the Bondi-Sachs formalism is the fact that the Bondi gauge is preserved under an infinite set of residual symmetries. A
version of these symmetries is called the generalised Bondi-Metzner-Sachs (BMS) symmetries. These symmetries are generated by supertranslations and arbitrary diffeomorphisms on the two-sphere  \cite{Barnich:2009se, Barnich:2010eb,Campiglia:2014yka,Campiglia:2015yka}. The Ward identities of these symmetries are identical to Weinberg soft graviton theorem \cite{He:2014laa} and to the sub-leading soft graviton theorem \cite{Campiglia:2014yka}; see \cite{Strominger:2017zoo} for more details and further references.  This has led to the suggestion that the generalised BMS group is a symmetry of the quantum gravity S-matrix, which in turn led to exploration of the subject in earnest. 

In parallel set of developments, related questions in the context of de Sitter spacetime have been pursued. The subject received renewed impetus after the work of  Ashtekar, Bonga, and Kesavan (ABK) \cite{Ashtekar:2014zfa, Ashtekar:2015lla,  Ashtekar:2015ooa, Ashtekar:2015lxa}, 
who derived the quadrupolar formula for gravitational radiation based on the decoupling of the linearised field equations in a generalised  harmonic gauge in de Sitter spacetime \cite{deVega:1998ia}.  These linearised solutions have been explored further by several authors \cite{Date:2015kma, Date:2016uzr, Bonga:2017dlx, Hoque:2018byx,  Dobkowski-Rylko:2022dva, Bonga:2023eml, Hoque:2017xop, Chu:2016qxp, Compere:2023ktn}, though it remained unclear how these solutions relate to gravitational perturbations in de Sitter studied in other approaches (see for example refs.~\cite{Mukohyama:2000ui, Anninos:2011jp, Bini:2011gg}).

Concurrently, there  has been detailed exploration of the  Bondi gauge and the associated asymptotic expansion in de Sitter \cite{He:2015wfa, Chrusciel:2016oux, Saw:2016isu, Saw:2017amv, Szabados:2018erf, Poole:2018koa, Compere:2019bua, Compere:2020lrt, Chrusciel:2020rlz, Chrusciel:2020rlz2,  Kolanowski:2020wfg,   Poole:2021avh, Geiller:2022vto, Bonga:2023eml, Geiller:2024amx}. In the Bondi-Sachs framework, Bonga, Bunster, and P\'erez (BBP) \cite{Bonga:2023eml} studied several aspects of gravitational radiation in de Sitter in both the linear and the non-linear theory. Working in the outgoing null coordinate system as used by Bondi and Sachs for asymptotically flat spacetimes, BBP introduced boundary conditions for asymptotically de Sitter  to allow for gravitational radiation.\footnote {These boundary conditions were first  introduced in refs.~\cite{Chrusciel:2020rlz, Chrusciel:2020rlz2} in the context of linearised perturbations around de Sitter background.}  They also discussed linearised gravitational waves in Bondi gauge. 
For the study of linearised solutions they focused on solving the homogeneous perturbation equations in a partial wave expansion. Given these developments, a natural question to ask is how the BBP linearised gravitational  solutions   are related to the linearised solutions of Ashtekar, Bonga, and Kesavan (ABK) \cite{deVega:1998ia, Ashtekar:2015lla,  Ashtekar:2015ooa, Ashtekar:2015lxa}. A look at these papers reveals that the question is far from trivial.

The Bondi-Sachs formalism famously clarified many subtleties in flat spacetime for the gravitational radiation theory. It is expected that the same would be the case for de Sitter spacetime. In particular, to address the question raised in the previous paragraph, a good exercise would be to map the ABK linearised solutions from the generalised harmonic gauge to Bondi gauge.\footnote{This exercise in flat spacetime has offered useful insight in the workings of the multipolar expansions \cite{Blanchet:2020ngx}.} This was done in a tour-de-force paper by Comp\`ere, Hoque, and Kutluk (CHK) \cite{Compere:2023ktn}. They found that there are additional subtleties in the  quadrupolar truncation considered by ABK.

Given the results of CHK, a more refined version of the question raised above can be asked: how the CHK linearised solutions are related to the BBP linearised  solutions? A main aim of the present paper is to answer this question. An additional complication in this analysis is the difference in the boundary conditions that BBP and CHK use. 
CHK use the boundary conditions introduced by Comp\`ere, Fiorucci, and Ruzziconi 
\cite{Compere:2019bua, Compere:2020lrt}, while BBP use the boundary conditions introduced in \cite{Chrusciel:2020rlz, Chrusciel:2020rlz2}. Both of these boundary conditions allow for  gravitational radiation in asymptotically de Sitter spacetimes.
A crucial difference for both Comp\`ere-Fiorucci-Ruzziconi and Bonga-Bunster-P\'erez boundary conditions compared to the asymptotically flat case is that the wave fields do not vanish at large distances. In fact, it is of the same order as the de Sitter space. In AdS/CFT parlance, gravitational radiation in de Sitter necessarily requires turning on non-normalisable modes in de Sitter spacetime. The fact that this must be the case has been emphasised by a number of authors; see for example discussions in \cite{Ashtekar:2015lla, Compere:2019bua}.  The difference in the boundary conditions used by CHK and BBP needs to be taken into account while relating these linearised  solutions. The difference is closely related to the discussion of residual symmetries preserving the Bondi gauge.

There are other missing links. Teukolsky \cite{Teukolsky:1982nz} in an earlier paper had written linearised quadrupolar solutions in flat spacetimes. How the BBP or the CHK linearised gravitational wave solutions relate to Teukolsky solutions in the flat spacetime limit? We also answer this question in this paper. In answering this question, we find unity in diversity, as we explain shortly. This is the reason we have chosen the title of our paper to be de Sitter Teukolsky waves. 

\begin{figure}
\centering
\begin{tikzpicture}[auto]
\node[entity] (node1) {de Sitter Teukolsky waves}
[grow=down, sibling distance=5cm]
child {node[attribute, below=1cm] {CHK waves \cite{Ashtekar:2015lxa, Compere:2023ktn}}}
child {node[attribute, below=1cm] {BBP waves \cite{Bonga:2023eml}}}
child {node[attribute, below=1cm] {LS waves \cite{Loganayagam:2023pfb, Loganayagam:2023pfb_2}}};
\end{tikzpicture}
\caption{\sl The organisation of this paper. In section \ref{sec:dS_Teukolsky}, we present the electric and magnetic parity de Sitter Teukolsky waves. In later sections, we relate de Sitter Teukolsky waves to wider literature. We exhibit gauge transformations starting from de Sitter Teukolsky waves. } \label{fig:organisation}
\end{figure}
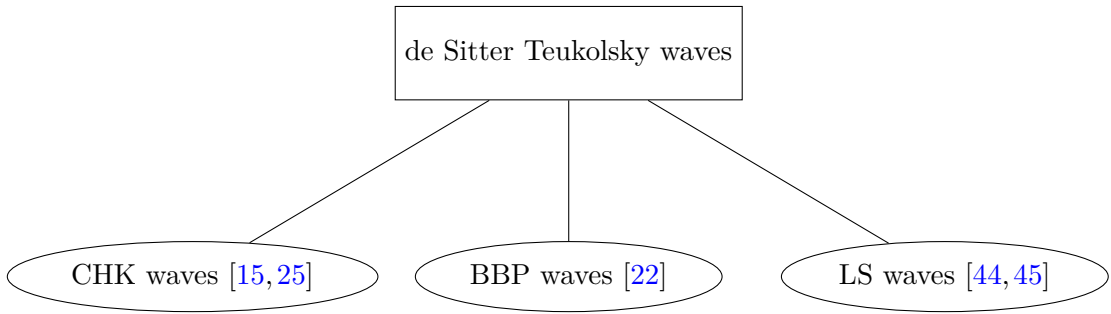

The rest of the paper is organised as follows. In section \ref{sec:dS_Teukolsky}, we present the electric and magnetic parity de Sitter Teukolsky waves. We present these solution in the transverse-traceless gauge. In the transverse-traceless gauge, linearised Einstein equations simplify significantly. As a result, without resorting to other detailed formalisms we are able to construct linearised quadrupolar solutions of interest straightforwardly.\footnote{The construction is explained in appendix \ref{app:solving_E_eqs}.}  Our presentation has the key advantage that in the cosmological constant going to zero limit the solutions reduce to Teukolsky solutions in flat spacetime \cite{Teukolsky:1982nz}.  In later sections, we relate de Sitter Teukolsky solutions to the wider literature. We exhibit gauge transformations \textit{starting from de Sitter Teukolsky solutions to other related solutions reported in the literature.}

In section \ref{sec:BBP}, we show, following closely the technology of Regge-Wheeler gauge fixing  \cite[Chapter 12]{Maggiore:2018sht}, that our linearised solutions are equivalent to the recently reported BBP linearised solutions in the Bondi gauge. In section \ref{sec:ls_construction}, we show that our linearised solutions are also equivalent  de Sitter gravitational perturbations studied by Loganayagam and Sheyte \cite{Loganayagam:2023pfb, Loganayagam:2023pfb_2} using a master variable formalism.  In section \ref{sec:CHK_dS}, we explore in detail the relation of our solutions  to the CHK solutions~\cite{Compere:2023ktn}. We show that our linearised solutions are also equivalent to their solutions, which can be thought of as an improvement over the ABK linearised solutions. The fact that our quadrupolar solutions precisely match with CHK solutions~\cite{Compere:2023ktn} is a highly non-trivial consistency check not only on our calculations and but also on the arguments made in \cite{Compere:2023ktn} in order to construct a consistent quadrupolar truncation in the ``post-de Sitter'' multipolar expansion. This section is more technical than the rest of the paper, partly because we need additional notation for relating asymptotic fields in Bondi gauge to source moments. For ease of comparison we closely follow the notation of \cite{Compere:2023ktn} in section  \ref{sec:CHK_dS}.  In section \ref{sec:energy}, we compute the energy flux  for de Sitter Teukolsky waves across future timelike infinity and find it to be manifestly positive. We end with a brief discussion of possible future directions in section \ref{sec:Conclusions}. 

Throughout the paper, we assume familiarity with the ten Zerilli tensor harmonics. We mostly follow the notation of  Maggiore \cite[Chapter 3]{Maggiore:2007ulw} \cite[Chapter 12]{Maggiore:2018sht}. Angular dependence formulae are collected in appendix \ref{sec:Zerilli}. As and when we need various Zerilli tensor harmonics, we also introduce them in the main text.

The organisation of this paper is also summarised in Figure~\ref{fig:organisation}.

\section{de Sitter Teukolsky waves}
\label{sec:dS_Teukolsky}
We work in  Bondi coordinates $\{u,r,\theta, \phi\}$, with de Sitter spacetime metric, 
\be
d\bar s^2 = \bar g_{\mu \nu} dx^\mu dx^\nu = -  \left(1- \frac{r^2}{L^2} \right) du^2 - 2 du dr + r^2 (d \theta^2 + \sin^2 \theta d\phi^2). \label{dS_metric}
\ee
This metric solves Einstein equations with positive cosmological constant $\Lambda$,
\be
R_{\mu \nu} - \frac{1}{2} R g_{\mu \nu} + \Lambda g_{\mu \nu} =0,
\ee
where the cosmological constant is related to the de Sitter length $L$ as,
\be 
\Lambda = \frac{3}{L^2}.
\ee
It is also convenient to define the Hubble parameter $H$ as the inverse of the de Sitter length $L$, $H = L^{-1}$.

  \begin{figure}[t]
\begin{center}
  \includegraphics[width=10cm]{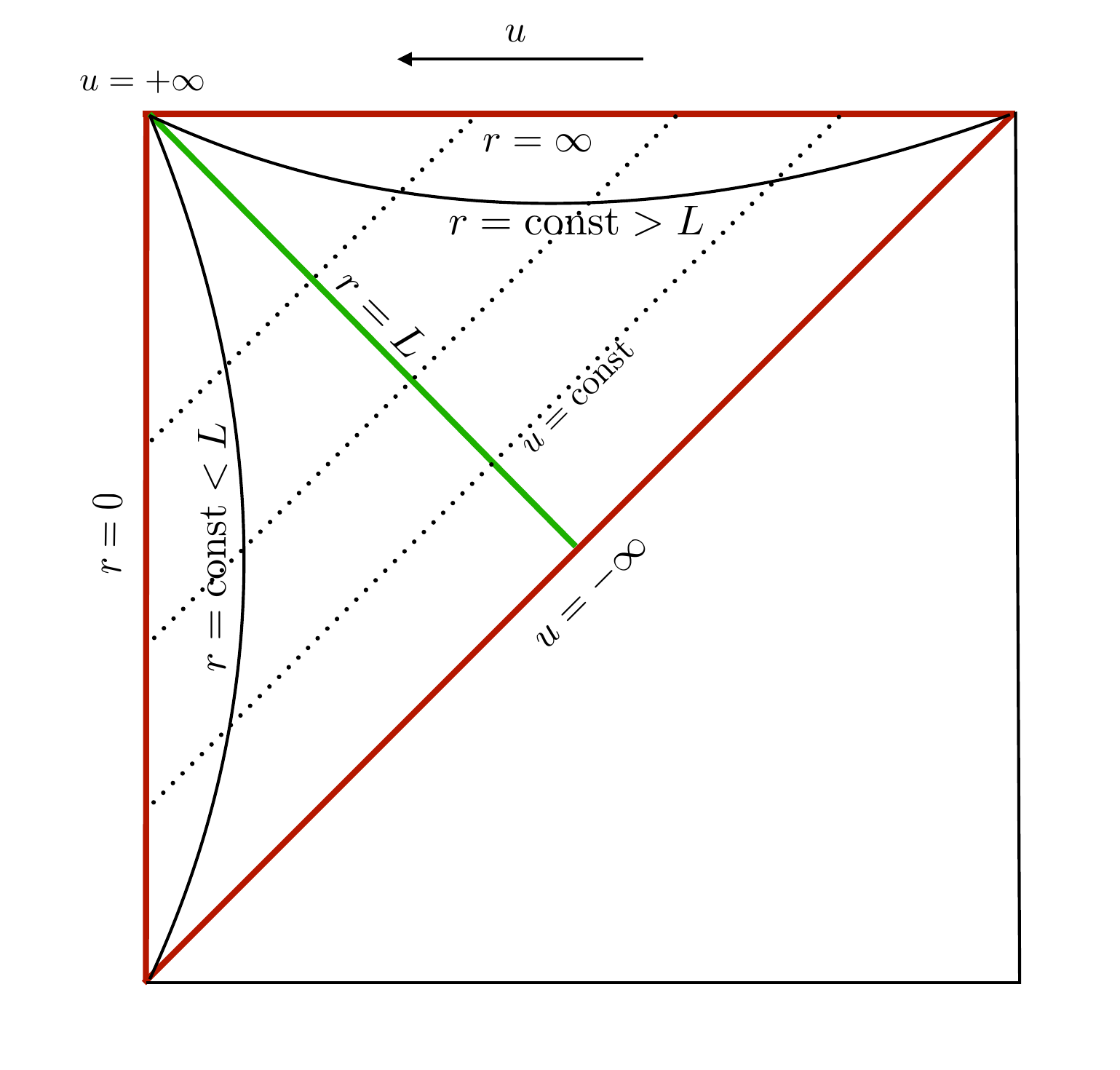}
\end{center}
\caption{\sl  Penrose diagram of de Sitter spacetime.  In Bondi coordinates $\{ u, r, \theta, \phi \}$, the future timelike infinity (a spacelike surface) is at $r \to \infty$. The dotted curves represent constant $u$  null hupersurfaces. The thin solid curves are constant $r$  surfaces. The nature of a constant $r$ surface depends on the value of $r$: for $r < L$  it is timelike; for $r > L$ it is spacelike. For $r=L$ the surface is null. The $r=L$ null surface serves as the cosmological horizon to the de Sitter static patch $r < L$.  The red triangle is called  the Poincar\'e patch. For most of this paper, we focus on the $r> L$ region. }
\label{fig:dS}
  \end{figure}

 Throughout the paper, we  mostly focus on the $r> L$ region in these coordinates. See Fig.~\ref{fig:dS}. In matrix form (which will be most useful in the following) de Sitter metric reads,
\be 
\bar g_{\mu \nu} = \left(
\begin{array}{cccc}
- \left(1 -  \frac{r^2}{L^2}\right) & -1 & 0 & 0 \\
-1 & 0 & 0 & 0 \\
0 & 0 & r^2 & 0 \\
0 & 0 & 0 & r^2 \sin ^2\theta  \\
\end{array}
\right), 
\qquad 
\bar g^{\mu \nu} =  
\left(
\begin{array}{cccc}
0 & -1 & 0 & 0 \\
-1 & \left(1-\frac{r^2}{L^2}\right) & 0 & 0 \\
0 & 0 & \frac{1}{r^2} & 0 \\
0 & 0 & 0 & \frac{1}{r^2 \sin ^2\theta } \\
\end{array}
\right). \label{dS_metric_matrix}
\ee
In this paper, we are interested in the linearised gravitational perturbations off de Sitter spacetime, 
\be
g_{\mu \nu} = \bar g_{\mu \nu} + \epsilon \, h_{\mu \nu}. 
\ee
The Einstein's equations for $g_{\mu \nu}$ expanded to first order in $\epsilon$ give the linearised Einstein equations for $h_{\mu \nu}$. In terms of the
trace reversed combination $\tilde{h}_{\mu\nu} := h_{\mu\nu} -
\tfrac{1}{2}h_{\mu\nu}(\bar{g}^{\alpha\beta}h_{\alpha\beta})$, the
linearised equations take the form (see, e.g., \cite{Date:2015kma}),
\begin{equation}
- \bar{\Box} \tilde{h}_{\mu\nu} + \left\{
\bar{\nabla}_{\mu}B_{\nu} + \bar{\nabla}_{\nu}B_{\mu} -
\bar{g}_{\mu\nu}(\bar{\nabla}^{\alpha}B_{\alpha})\right\} +
\frac{2\Lambda}{3}\left(\tilde{h}_{\mu\nu} - \tilde{h}\bar{g}_{\mu\nu}\right) 
= 0, \label{LinEqn}
\end{equation}
where $B_{\mu} := \bar{\nabla}_{\alpha}\tilde{h}^{\alpha}_{~\mu}$,  $\bar{\Box} = \bar \nabla_\alpha \bar \nabla^\alpha$, 
and where $\bar \nabla$ is the covariant derivative with respect to de Sitter background metric. 
In the transverse traceless gauge,
$\bar{\nabla}^{\mu} \tilde{h}_{\mu \nu}=0, 
\tilde{h} = \bar g^{\mu \nu} \tilde{h}_{\mu \nu} = 0,$
which implies
\begin{align}
\bar{\nabla}^{\mu} h_{\mu \nu}&=0, &
h = \bar g^{\mu \nu} h_{\mu \nu} &= 0,
\label{transverse_condition}
\end{align}
and as a result, linearised Einstein's equations \eqref{LinEqn} on de Sitter background become 
\be
\left({\bar \nabla}^\rho {\bar \nabla}_\rho - \frac{2}{L^2 }\right) h_{\mu \nu} =0. \label{Einstein_equation}
\ee

\subsection{Magnetic parity}
\label{sec:magnetic_Teukolsky}
A close look at the magnetic parity Teukolsky waves in flat spacetime \cite{Teukolsky:1982nz} reveals that they are based on a set of two tensor harmonics, $\mathbb{T}^\mathrm{B1}$ and $\mathbb{T}^\mathrm{B2}$. Magnetic parity Bonga-Bunster-P\'erez (BBP) solutions \cite{Bonga:2023eml} in \textit{Bondi-Sachs gauge} in de Sitter are also based on a set of two tensor harmonics, but this set is different from that used by Teukolsky. BBP solutions are based on $\mathbb{T}^\mathrm{Bu}$ and $\mathbb{T}^\mathrm{B2}$. 
Given this difference, it is natural to construct Teukolsky waves in de Sitter using all three magnetic parity tensor harmonics:
\be
\mathbb{T}^\mathrm{Bu}, \quad \mathbb{T}^\mathrm{B1}, \quad \mathbb{T}^\mathrm{B2}. \label{three_tensor_harmonics}
\ee
For ready reference, we list the three magnetic parity tensor harmonics in Table~\ref{table:magnetic}. More details are in appendix \ref{sec:Zerilli}.	
\begin{table}[t!]
\centering
\begin{tabular}{||c|c||} 
\hline 
$(\mathbb{T}^\mathrm{Bu}_{lm} )_{\mu \nu} $
&
{\small
$
\left(
\begin{array}{cccc}
0 & 0 & \frac{1}{\sin \theta} \partial_\phi Y_{lm}& - \sin \theta \partial_\theta Y_{lm} \\
0 & 0 & 0 & 0 \\
\frac{1}{\sin \theta} \partial_\phi Y_{lm} & 0 & 0 & 0 \\
- \sin \theta \partial_\theta Y_{lm} & 0 & 0 & 0\\
\end{array}
\right)
$
}
\\
\hline
$
(\mathbb{T}^\mathrm{B1}_{lm} )_{\mu \nu} 
$
&
{\small 
$
\left(
\begin{array}{cccc}
0 & 0 & 0 & 0 \\
0 & 0 & \frac{1}{\sin \theta} \partial_\phi Y_{lm}& - \sin \theta \partial_\theta Y_{lm} \\
0 &  \frac{1}{\sin \theta} \partial_\phi Y_{lm} & 0 & 0 \\
0 & - \sin \theta \partial_\theta Y_{lm} & 0 & 0\\
\end{array}
\right) $  
}
\\
\hline
$(\mathbb{T}^\mathrm{B2}_{lm} )_{\mu \nu} 
$
& 
{\small 
$
\left(
\begin{array}{cccc}
0 & 0 & 0 & 0 \\
0 & 0 & 0 & 0 \\
0 & 0 & -\frac{1}{\sin \theta}  \,X( Y_{lm})& \sin \theta \, W( Y_{lm}) \\
0 & 0 & \sin \theta  \, W( Y_{lm})&\sin \theta  \, X( Y_{lm}) \\
\end{array}
\right)
$ 
}\\
\hline
\end{tabular}
\caption{\sl Three magnetic parity tensor harmonics. The matrices are in coordinates $\{ u, r, \theta, \phi \}$.  For more details refer to appendix \ref{sec:Zerilli}. }
\label{table:magnetic}
\end{table}

We consider the general linear combination of all three magnetic tensor harmonics, 
\be
h_{\mu \nu} = \sum_{m=-l}^{l} \left(f_{lm}^\mathrm{Bu}(u,r) (\mathbb{T}^\mathrm{Bu}_{lm} )_{\mu \nu}+  f_{lm}^\mathrm{B1}(u,r) (\mathbb{T}^\mathrm{B1}_{lm} )_{\mu \nu}+  f_{lm}^\mathrm{B2}(u,r) (\mathbb{T}^\mathrm{B2}_{lm} )_{\mu \nu} \right)\label{magnetic_teukolsky},
\ee
with $l=2$, where $ f_{2m}^\mathrm{Bu}(u,r),  f_{2m}^\mathrm{B1}(u,r),  f_{2m}^\mathrm{B2}(u,r)$ are functions to be determined. Note that since the three tensor harmonics \eqref{three_tensor_harmonics} are traceless, $h_{\mu \nu}$ is traceless as well.  The functions 
\bea
f_{2m}^\mathrm{B1} &=& \frac{4}{r} \ddot{B}_{m} + \frac{12}{r^2} \left(1+ \frac{r^2}{3L^2}\right) \dot{B}_{m}+ \frac{12}{r^3}\left(1 + \frac{r^2}{3L^2}  \right) B_m , \label{coeB1}\\
f_{2m}^\mathrm{B2} &=& r B_m^{(3)} + 2\left( 1 + \frac{r^2}{2L^2} \right) \ddot{B}_{m} + \frac{3}{r} \left(1+ \frac{r^2}{L^2}\right) \dot{B}_{m}+ \frac{3}{r^2}\left(1 + \frac{2r^2}{3L^2} + \frac{r^4}{L^4}  \right) B_m, \label{coeB2}\\
f_{2m}^\mathrm{Bu} &=& - \frac{4}{L^2} \left[ r \ddot{B}_{m}+ \left( 1 + \frac{r^2}{L^2} \right) \dot{B}_{m}\right], \label{coeBu}
\eea
solve the transversality condition \eqref{transverse_condition} and linearised Einstein equations \eqref{Einstein_equation} for arbitrary functions $B_m(u)$. Here $\dot{B}_m$ and $\ddot{B}_m$  denote respectively  the first and the second derivative of the function $B_m(u)$ with respect to $u$.  $ B_m^{(k)}$ denotes $k$-th derivative with respect to $u$.  
In the $L \to \infty$ limit, the solutions reduce to the magnetic parity Teukolsky solutions~\cite{Teukolsky:1982nz},
\bea \label{mag_T_flat_1}
f_{2m}^\mathrm{B1} &=& \frac{4 \ddot{B}_m}{r}   + \frac{12  \dot{B}_{m} }{r^2} + \frac{12 B_m}{r^3}  ,\\
f_{2m}^\mathrm{B2} &=& r B_m^{(3)} +2 \ddot{B}_{m} + \frac{3\dot{B}_{m}}{r}+ \frac{3B_m}{r^2},\\
f_{2m}^\mathrm{Bu} &=& 0. \label{mag_T_flat_3}
\eea
Some comments on how we constructed this solution are presented in appendix \ref{app:solving_E_eqs}. Some further aspects of the original Teukolsky solutions~\cite{Teukolsky:1982nz} are studied in appendix \ref{app:CC_zero}.

\subsection{Electric parity}

\label{sec:electric_Teukolsky}

For the electric parity Teukolsky waves in flat spacetime we need three tensor harmonics, $\mathbb{T}^\mathrm{S0}$, $\mathbb{T}^\mathrm{E1}$ and $\mathbb{T}^\mathrm{E2}$.  For BBP solutions in Bondi-Sachs gauge in de Sitter spacetime
we also need three tensor harmonics, but they are not the same as the one used by Teukolsky.  BBP  use $\mathbb{T}^\mathrm{uu}$,  $\mathbb{T}^\mathrm{Eu}$, and $\mathbb{T}^\mathrm{E2}$.  Therefore, it is natural to guess that to construct Teukolsky waves in de Sitter we need all six traceless tensor harmonics. 
At this stage it is easiest to work with the following set:
\be
\mathbb{T}^\mathrm{uu}, \quad \mathbb{T}^\mathrm{ur}, \quad \mathbb{T}^\mathrm{Eu}, \quad \mathbb{T}^\mathrm{S0}, \quad \mathbb{T}^\mathrm{E1},  \quad \mathbb{T}^\mathrm{E2}.
\label{six_tensor_harmonics_electric}
\ee
For ready reference we list these six traceless electric parity tensor harmonics in Table~\ref{table:electric}. For more details, we refer the reader to appendix \ref{sec:Zerilli}. 

%
%
\begin{table}[t!]
\centering
\begin{tabular}{||c|c||} 
\hline 
$
(\mathbb{T}^\mathrm{uu}_{lm} )_{\mu \nu}
$
&
{\small 
$
\left(
\begin{array}{cccc}
1& 0 & 0& 0 \\
0 & 0 & 0 & 0 \\
0 & 0 & 0 & 0 \\
0 & 0 & 0 & 0 \\
\end{array}
\right) Y_{lm}
$ 
} 
\\
\hline 
$
(\mathbb{T}^\mathrm{ur}_{lm} )_{\mu \nu} 
$ 
&
{\small 
$
\left(
\begin{array}{cccc}
0& 1 & 0& 0 \\
1 &2 \left(1-\frac{r^2}{L^2}\right)^{-1} & 0 & 0 \\
0 & 0 & 0 & 0 \\
0 & 0 & 0 & 0 \\
\end{array}
\right) Y_{lm}
$
} \\
\hline
$
(\mathbb{T}^\mathrm{Eu}_{lm} )_{\mu \nu}
$
&
{\small 
$
\left(
\begin{array}{cccc}
0 & 0 & \partial_\theta Y_{lm}&  \partial_\phi Y_{lm} \\
0 & 0 & 0 & 0 \\
\partial_\theta Y_{lm} & 0 & 0 & 0\\
\partial_\phi Y_{lm}& 0 & 0 & 0\\
\end{array}
\right)
$
} \\
\hline
$
(\mathbb{T}^\mathrm{S0}_{lm} )_{\mu \nu}
$
&
{\small 
$
\left(
\begin{array}{cccc}
0& 0 & 0& 0 \\
0 & \left(1-\frac{r^2}{L^2}\right)^{-1} & 0 & 0 \\
0 & 0 & -\frac{1}{2} r^2 & 0 \\
0 & 0 & 0 & -\frac{1}{2} r^2 \sin^2 \theta \\
\end{array}
\right) Y_{lm}
$
}
\\
\hline
$
(\mathbb{T}^\mathrm{E1}_{lm} )_{\mu \nu}
$
&
{\small 
$
\left(
\begin{array}{cccc}
0 & 0 & 0 & 0 \\
0 & 0 & \partial_\theta Y_{lm}&  \partial_\phi Y_{lm} \\
0 & \partial_\theta Y_{lm} & 0 & 0 \\
0 &\partial_\phi Y_{lm}& 0 & 0\\
\end{array}
\right)
$
}
\\
\hline
$
(\mathbb{T}^\mathrm{E2}_{lm} )_{\mu \nu}
$
&
{\small 
$
\left(
\begin{array}{cccc}
0 & 0 & 0 & 0 \\
0 & 0 & 0 & 0 \\
0 & 0 &W( Y_{lm})& X(Y_{lm}) \,  \\
0 & 0 & X(Y_{lm})& - \sin^2 \theta \, W( Y_{lm}) \\
\end{array}
\right)
$ 
} \\
\hline
\end{tabular}
\caption{\sl Six traceless electric parity tensor harmonics. The matrices are in coordinates $\{ u, r, \theta, \phi \}$. The trace is with respect to the background metric $\bar g_{\mu \nu}$ \eqref{dS_metric_matrix}: $h =  \bar g^{\mu \nu} h_{\mu \nu}$.   For more details refer to appendix \ref{sec:Zerilli}. }
\label{table:electric}
\end{table}


The linear combination,
\bea \label{electric_teukolsky}
h_{\mu \nu} &=& \sum_{m=-2}^{2} \left( A_m^\mathrm{uu} (u,r) (\mathbb{T}^\mathrm{uu}_{2m} )_{\mu \nu}+ A_m^\mathrm{ur} ( \mathbb{T}^\mathrm{ur}_{2m})_{\mu \nu} + A_m^\mathrm{Eu}(u,r) (\mathbb{T}^\mathrm{Eu}_{2m} )_{\mu \nu} \right. \nonumber  \\ 
& &~ \left. + A_m^\mathrm{S0} (u,r) (\mathbb{T}^\mathrm{S0}_{2m} )_{\mu \nu}+A_m^\mathrm{E1}(u,r) (\mathbb{T}^\mathrm{E1}_{2m} )_{\mu \nu}+ A_m^\mathrm{E2}(u,r) (\mathbb{T}^\mathrm{E2}_{2m} )_{\mu \nu} \right),
\eea
with the functions, 
\bea
\label{Amuu}
A_m^\mathrm{uu} &=& \frac{1}{L^2} \left[ 
-\frac{1}{r} \left(1-\frac{r^2}{L^2} \right) \ddot{A}_{m}
-\frac{3}{r^2}\left(1-\frac{2 r^2}{3 L^2}-\frac{r^4}{3 L^4}\right)\dot{A}_{m}  
-\frac{3 }{r^3}\left(1-\frac{r^2}{L^2}\right)A_m \right], \\
A_m^\mathrm{ur} &=& \frac{1}{L^2} \left[ -\frac{1}{
r}\ddot{A}_{m} -\frac{3}{r^2} \left(1+\frac{r^2}{3 L^2}\right) \dot{A}_{m} -\frac{3}{r^3}A_m \right],\\
A_m^\mathrm{Eu} &=& \frac{1}{L^2} \left[\frac{r}{6} A_m^{(3)}+ \frac{1}{2} \left(1+\frac{r^2}{3L^2} \right) \ddot{A}_{m} + \frac{1}{2r} \dot{A}_{m} \right],\\
A_m^\mathrm{S0} &=&\frac{1}{r^3} \left(1+\frac{r^2}{L^2} \right) \ddot{A}_{m}+ \frac{3}{r^4} \left(1+\frac{4 r^2}{3 L^2}+\frac{r^4}{3 L^4}\right)\dot{A}_{m} + \frac{3}{r^5} \left(1+ \frac{r^2}{L^2} \right)A_m ,\\
A_m^\mathrm{E1} &=& -\frac{A_m^{(3)}}{6 r}-\frac{1}{2 r^2} \left( 1+ \frac{r^2}{3L^2}\right) \ddot{A}_{m}-\frac{1}{r^3}\left( 1- \frac{r^2}{6L^2}\right) \dot{A}_{m}-\frac{A_m}{r^4},\\
A_m^\mathrm{E2} &=& \frac{1}{24} r A_m^{(4)} + \frac{1}{12} \left(1+ \frac{r^2}{2 L^2}\right) A_m^{(3)} + \frac{1}{8 r} \left(1-\frac{2 r^2}{L^2} \right) \ddot{A}_{m} + \frac{1}{8 r^2} \left(1-\frac{8 r^2}{3 L^2}-\frac{2 r^4}{L^4}\right) \dot{A}_{m}  \nonumber \\  &&  +  \frac{1}{8 r^3}\left(1-\frac{3 r^2}{L^2} \right)A_m \label{AmE2}
\eea
solve the transversality condition \eqref{transverse_condition} and linearised Einstein equations \eqref{Einstein_equation} for arbitrary functions $A_m(u)$.
As before,  $\dot{A}_m$ and $\ddot{A}_m$  denote respectively the first and the second derivative of the function $A_m(u)$ with respect to $u$ and $A_m^{(k)}$ denotes $k$-th derivative with respect to $u$.  In the $L \to \infty$ limit, the solutions reduce to the electric parity Teukolsky solutions~\cite{Teukolsky:1982nz},
\begin{align} \label{elec_T_flat_1}
A_m^\mathrm{uu} &=   0, \\ 
A_m^\mathrm{ur} &=0,\\
A_m^\mathrm{Eu}&=0, \\
A_m^\mathrm{S0} &= \frac{\ddot{A}_{m}}{r^3}+\frac{3 \dot{A}_{m}}{r^4}+\frac{3 A_m}{r^5}, \\
A_m^\mathrm{E1} &= -\frac{A_m^{(3)}}{6 r}-\frac{\ddot{A}_{m}}{2 r^2}-\frac{\dot{A}_{m}}{r^3}-\frac{A_m}{r^4}, \\
A_m^\mathrm{E2} &=  \frac{A_m^{(4)}}{24} r +\frac{A_m^{(3)}}{12} +\frac{\ddot{A}_{m}}{8 r}+\frac{\dot{A}_{m}}{8
r^2}+\frac{A_m}{8 r^3}. \label{elec_T_flat_3}
\end{align}

\section{Relation to Bonga-Bunster-P\'erez waves}

\label{sec:BBP}

Bonga, Bunster, and P\'erez \cite{Bonga:2023eml} studied several aspects of gravitational radiation in de Sitter spacetime in both the linear and the non-linear theory. Working in the null coordinate system as used by Bondi and Sachs for asymptotically flat case, they introduced boundary conditions for asymptotically de Sitter  to allow for gravitational radiation.

In the Bondi-Sachs framework adapted to asymptotically de Sitter context, BBP  also studied explicit solutions to the linearised Einstein's equation. They focused on solving the homogeneous perturbation equations in a partial wave expansion. Since the background de Sitter spacetime is spherically symmetric, they used techniques similar to Regge and Wheeler \cite{Regge:1957td}\footnote{For a concise review of the Regge-Wheeler gauge fixing we refer the reader to \cite[Chapter 12]{Maggiore:2018sht}.} (more precisely, they used techniques from \cite{Martel:2005ir}) but did not impose the Regge-Wheeler gauge.
In particular, they also introduced the scalar, vector, and tensor harmonics to separate the angular part from the radial part of the perturbations. In such a spherical harmonic decomposition, they constructed magnetic and electric parity solutions with $l=2$ tensor harmonics. 
We call these solutions the BBP waves.

Let $Y_{l m}(x^{A})$ be the usual scalar spherical-harmonic functions. Following \cite{Martel:2005ir}, let the even-parity (also known as electric) vector harmonics be $Y_{A}^{l m}$ and odd-parity (also known as magnetic) vector harmonics be  $X_{A}^{l m}$. They are  related to the scalar harmonics through the covariant derivative operator compatible with the round metric $\gamma_{AB}$ on the unit two-sphere ($\gamma_{AB} dx^A dx^B = d\theta^2 + \sin^2 \theta d\phi^2$)  as,
\begin{align}
Y_{A}^{l m} & =D_{A}Y_{l m}, &
X_{A}^{l m} & = - \epsilon_{A}{}^{B}D_{B}Y_{l m}.
\end{align}
Similarly, the electric and magnetic tensor harmonics are, 
\begin{align}
Y_{AB}^{l m} & :=D_{(A}Y_{B)}^{l m}-\frac{1}{2}\gamma_{AB}D_{C}Y_{l m}^{C}, &
X_{AB}^{l m} & :=-\epsilon_{(A}{}^{C}D_{B)}Y_{C}^{l m},
\end{align}
where for an arbitrary tensor $T_{AB}$, $T_{(AB)} = \tfrac{1}{2}(T_{AB} + T_{BA})$, and $\epsilon_{AB}$ is the Levi-Civita tensor on the unit two-sphere, $\epsilon_{\theta \phi} = \sin \theta$.

In terms of these harmonics, the magnetic\footnote{We have corrected the typo of an overall minus sign in the $h_{uA}^\text{(B)}$ term in \cite{Bonga:2023eml}.} (odd) and electric (even) parity BBP waves take the form,
\begin{align}
\label{eq:magsol1} 
h_{uu}^{\text{(B)}} & =0,\\
h_{ur}^{\text{(B)}} & =0,\\
h_{uA}^{\text{(B)}} & = - \sum_{m=-2}^{2}\left[\frac{1}{2}\left(\ddot{b}_{m}- \frac{1}{L^2}b_{m}\right)\frac{r^{2}}{L^{2}}+\left(\ddot{b}_{m} - \frac{1}{L^2}b_{m}\right)
+\frac{2}{r}\dot{b}_{m}+\frac{3}{2r^{2}}b_{m}\right]X_{A}^{2m},\\
h_{AB}^{\text{(B)}} & =\sum_{m=-2}^{2}\left[r\left(\ddot{b}_{m}-\frac{1}{L^2}b_{m}\right)-\frac{1}{r}b_{m}\right]X_{AB}^{2m},
\label{eq:magsol2} 
\end{align}
and	
\begin{align}
\label{eq:elecsol1} 
h_{uu}^{\text{(E)}} & =\sum_{m=-2}^{2}\left[3\left(\ddot{a}_{m}- \frac{4}{L^2}a_{m}\right)\frac{r}{L^{2}}
+6\left(\ddot{a}_{m}-\frac{1}{L^2}a_{m}\right)\frac{1}{r}+\frac{6}{r^2}\dot{a}_{m}+\frac{3}{r^{3}}a_{m}\right]Y_{2m},\\
h_{ur}^{\text{(E)}} & =0,\\
h_{uA}^{\text{(E)}} & =\sum_{m=-2}^{2}\left[\left(\frac{2}{L^2}a_{m}-\frac{1}{2}\ddot{a}_{m}\right)\frac{r^2}{L^2}+\left(\frac{4}{L^2}a_{m}-\ddot{a}_{m}\right)
+\frac{2}{r}\dot{a}_{m}+\frac{3}{2r^{2}}a_{m}\right]Y_{A}^{2m},\\
h_{AB}^{\text{(E)}} & =\sum_{m=-2}^{2}\left[r\left(\ddot{a}_{m}-\frac{4}{L^{2}}a_{m}\right)+\frac{1}{r}a_{m}\right]Y_{AB}^{2m},
\label{eq:elecsol2} 
\end{align}
respectively.  In these equations, $a_m$ and $b_m$ are arbitrary functions of $u$ for each integer $m$ between $-2$ to $2$, and dots denote derivatives with respect to $u$. 
In the notation of tensor harmonics introduced in section \ref{sec:dS_Teukolsky}, the BBP magnetic parity solution \eqref{eq:magsol1}--\eqref{eq:magsol2} is
\bea
h_{\mu \nu}^{\text{(B)}}  &=&  \sum_{m=-2}^{2} \frac{1}{2}\left[r\left(\ddot{b}_{m}-\frac{1}{L^2}b_{m}\right)-\frac{1}{r}b_{m}\right](\mathbb{T}^{\mathrm{B2}}_{2m})_{\mu \nu} \nonumber \\ 
& & + 
\sum_{m=-2}^{2}
\left[\frac{1}{2}\left(\ddot{b}_{m}- \frac{1}{L^2}b_{m}\right)\frac{r^{2}}{L^{2}}+\left(\ddot{b}_{m} - \frac{1}{L^2}b_{m}\right)
+\frac{2}{r}\dot{b}_{m}+\frac{3}{2r^{2}}b_{m}\right] (\mathbb{T}^{\mathrm{Bu}}_{2m})_{\mu \nu} . \label{28III24.01_BBP}
\eea
In our conventions, 
\begin{align}
(\mathbb{T}_{lm}^\mathrm{Bu})_{uA} &= -X_{A}^{lm}, & 
(\mathbb{T}_{lm}^\mathrm{B2})_{AB} &= 2X_{AB}^{lm}.
\end{align}
The BBP electric parity solution \eqref{eq:elecsol1}--\eqref{eq:elecsol2}
is,
\begin{align} 
h_{\mu\nu}^{\text{(E)}}  & =\sum_{m=-2}^{2}\left[3\left(\ddot{a}_{m}- \frac{4}{L^2}a_{m}\right)\frac{r}{L^{2}}
+6\left(\ddot{a}_{m}-\frac{1}{L^2}a_{m}\right)\frac{1}{r}+\frac{6}{r^2}\dot{a}_{m}+\frac{3}{r^{3}}a_{m}\right](\mathbb{T}^\mathrm{uu}_{2m} )_{\mu \nu} \nonumber\\
&
+\sum_{m=-2}^{2}\left[\left(\frac{2}{L^2}a_{m}-\frac{1}{2}\ddot{a}_{m}\right)\frac{r^2}{L^2}+\left(\frac{4}{L^2}a_{m}-\ddot{a}_{m}\right)
+\frac{2}{r}\dot{a}_{m}+\frac{3}{2r^{2}}a_{m}\right](\mathbb{T}^\mathrm{Eu}_{2m} )_{\mu \nu}\nonumber\\
& +\sum_{m=-2}^{2}\frac{1}{2}\left[r\left(\ddot{a}_{m}-\frac{4}{L^{2}}a_{m}\right)+\frac{1}{r}a_{m}\right](\mathbb{T}^\mathrm{E2}_{2m} )_{\mu \nu},
\label{eq:elecsol_BBP}
\end{align}
where in our conventions, 
\begin{align}
(\mathbb{T}^\mathrm{uu}_{2m} )_{uu}&= Y^{lm} &
(\mathbb{T}_{lm}^\mathrm{Eu})_{uA} &= Y_{A}^{lm}, &
(\mathbb{T}_{lm}^\mathrm{E2})_{AB} &= 2Y_{AB}^{lm}.
\end{align}
We now show that via an infinitesimal diffeomorphism, 
\be
h_{\mu \nu}'(x) = h_{\mu \nu}(x) - \left(\bar{\nabla}_\mu \xi_\nu + \bar{\nabla}_\nu \xi_\mu \right),\label{diffeo}
\ee
we can relate the magnetic and electric parity de Sitter Teukolsky waves to BBP waves.

Observe that a general vector field $\xi_\mu(x)$ can be expanded in scalar and vector spherical harmonics as follows \cite[Chapter 12]{Maggiore:2018sht}
\bea
\xi_u (x) &=& \sum_{l=0}^{\infty} \sum_{m=-l}^{l} \xi^{(u)}_{lm} (u,r) Y_{lm}(x^A), \\
\xi_r (x) &=& \sum_{l=0}^{\infty} \sum_{m=-l}^{l} \xi^{\mathrm{(R)}}_{lm} (u,r) Y_{lm}(x^A),  \\
\xi_A (x) &=& \sum_{l=1}^{\infty} \sum_{m=-l}^{l} \xi^{\mathrm{(E)}}_{lm} (u,r)  Y_A^{lm}(x^B)  + \sum_{l=1}^{\infty} \sum_{m=-l}^{l} \xi^{\mathrm{(B)}}_{lm} (u,r) X_A^{lm}(x^B). 
\eea
The functions $\xi^{(u)}_{lm}, \xi^{\mathrm{(R)}}_{lm},$ and $\xi^{\mathrm{(E)}}_{lm}$ generate electric parity terms whereas the function $\xi^{\mathrm{(B)}}_{lm}$ generates magnetic parity terms. 
It is convenient to consider the magnetic and the electric parity solutions separately. We consider these transformations on the de Sitter Teukolsky waves \eqref{magnetic_teukolsky} and \eqref{electric_teukolsky}. 

\subsection{Magnetic parity}
We first consider the magnetic parity gauge transformations  \eqref{diffeo} for $l=2$. We have, 
\bea \label{mag_diffeo_1}
\xi_u (x) &=& 0, \\ 
\xi_r (x) &=&0,  \\
\xi_A (x) &=& \sum_{m=-2}^{2} \xi^{\mathrm{(B)}}_{2m} (u,r) X_A^{lm}(x^B).  \label{mag_diffeo_3}
\eea
We can now easily compute covariant derivatives needed to implement \eqref{diffeo}. In the basis of tensor harmonics given in appendix \ref{sec:Zerilli}, we get
\be
\left(\bar{\nabla}_\mu \xi_\nu + \bar{\nabla}_\nu \xi_\mu \right) = \sum_{m=-2}^{2} \left[  \xi^{\mathrm{(B)}}_{2m} (\mathbb{T}^{\mathrm{B2}}_{2m})_{\mu \nu} - \frac{1}{r} \left( r \partial_r \xi^{\mathrm{(B)}}_{2m} - 2 \xi^{\mathrm{(B)}}_{2m} \right) (\mathbb{T}^{\mathrm{B1}}_{2m})_{\mu \nu}- \partial_u \xi^{\mathrm{(B)}}_{2m} (\mathbb{T}^{\mathrm{Bu}}_{2m})_{\mu \nu} \right]. \label{covD_symmetrized}
\ee
From this equation, we see that the coefficients of various magnetic parity tensor harmonics transform as
\bea
\label{transform_magnetic_1}
f_{2m}^{\mathrm{B2}} &\to& f_{2m}^{\mathrm{B2}} - \xi^{\mathrm{(B)}}_{2m}, \\
\label{transform_magnetic_2}
f_{2m}^{\mathrm{B1}} &\to& f_{2m}^{\mathrm{B1}} + \frac{1}{r} \left( r \partial_r \xi^{\mathrm{(B)}}_{2m} - 2 \xi^{\mathrm{(B)}}_{2m} \right), \\
f_{2m}^{\mathrm{Bu}} &\to& f_{2m}^{\mathrm{Bu}} +  \partial_u \xi^{\mathrm{(B)}}_{2m}.
\label{transform_magnetic_3}
\eea

In particular, starting with the magnetic parity de Sitter Teukolsky waves \eqref{magnetic_teukolsky} 
we can get rid of the coefficient of $ (\mathbb{T}^{\mathrm{B1}}_{2m})_{\mu \nu}$ by appropriately choosing $\xi^{\mathrm{(B)}}_{2m} (u,r)$. Note that the coefficient of $ (\mathbb{T}^{\mathrm{B1}}_{2m})_{\mu \nu}$ is zero in the BBP magnetic parity solution \eqref{28III24.01}.  Setting the coefficient of $(\mathbb{T}^{\mathrm{B1}}_{2m})_{\mu \nu}$ to zero we have
\be
\frac{1}{r} \left( r \partial_r \xi^{\mathrm{(B)}}_{2m} - 2 \xi^{\mathrm{(B)}}_{2m} \right) +  \frac{4}{r} \ddot{B}_{m} + \frac{12}{r^2} \left(1+ \frac{r^2}{3L^2}\right) \dot{B}_{m}+ \frac{12}{r^3}\left(1 + \frac{r^2}{3L^2}  \right) B_m=0.
\ee
This is a differential equation in the radial coordinate. Integration of this equation gives, 
\be
\xi^{\mathrm{(B)}}_{2m}  = 2 \ddot{B}_m(u)  + \frac{4}{r} \left( 1 + \frac{r^2}{L^2}\right)  \dot{B}_m(u) + \frac{3}{r^2} \left( 1 + \frac{2r^2}{3L^2}\right)  B_m(u) + r^2 C_m(u),
\ee
where $C_m(u)$ is an integration `constant'.  The choice
\be
C_m(u) = \frac{3}{L^4} B_m(u) + \frac{1}{L^2} \ddot{B}_m(u),
\ee
with the identification 
\be
B_m(u) = \frac{1}{2}\int^u b_m(u') du' , \label{iden-BBP_mag}
\ee
maps the magnetic parity de Sitter Teukolsky waves to the magnetic parity BBP waves.

\subsection{Electric parity}
\label{sec:EP_BBP}
For the electric parity case, the discussion is almost identical, but a little more involved. 
For the implementation of the general electric parity diffeomorphism \eqref{diffeo}, it is more convenient to work with a set of  tensor harmonics slightly different from the one used in writing the de Sitter Teukolsky solution \eqref{electric_teukolsky}. We introduce,
\be
(\mathbb{T}^\mathrm{Ru}_{lm} )_{\mu \nu} =\left(
\begin{array}{cccc}
0& 1 & 0& 0 \\
1 & 0 & 0 & 0 \\
0 & 0 & 0 & 0 \\
0 & 0 & 0 & 0 \\
\end{array}
\right) Y_{lm},
\quad \quad
(\mathbb{T}^\mathrm{L0}_{lm} )_{\mu \nu} =\left(
\begin{array}{cccc}
0& 0 & 0& 0 \\
0 & 1 & 0 & 0 \\
0 & 0 &0 & 0 \\
0 & 0 & 0 &0 \\
\end{array}
\right) Y_{lm},
\ee
\be
(\mathbb{T}^\mathrm{T0}_{lm} )_{\mu \nu} =\left(
\begin{array}{cccc}
0& 0 & 0& 0 \\
0 & 0 & 0 & 0 \\
0 & 0 &1 & 0 \\
0 & 0 & 0 &\sin^2 \theta \\
\end{array}
\right) Y_{lm},
\ee
For more details, we  refer the reader to appendix \ref{sec:Zerilli}. 
The seven electric parity tensor harmonics are,
\be
\mathbb{T}^\mathrm{uu}, \quad \mathbb{T}^\mathrm{Ru}, \quad \mathbb{T}^\mathrm{Eu}, \quad \mathbb{T}^\mathrm{L0}, \quad \mathbb{T}^\mathrm{T0}, \quad \mathbb{T}^\mathrm{E1},  \quad \mathbb{T}^\mathrm{E2}.
\label{seven_tensor_harmonics_electric}
\ee
In this set, we have essentially replaced $\mathbb{T}^\mathrm{S0}$ and $ \mathbb{T}^\mathrm{ur}$ in favour of three tensor harmonics $\mathbb{T}^\mathrm{Ru}, \mathbb{T}^\mathrm{L0},$ and $\mathbb{T}^\mathrm{T0}$. The tracefree combinations $\mathbb{T}^\mathrm{S0}$ and $ \mathbb{T}^\mathrm{ur}$ are related to $\mathbb{T}^\mathrm{Ru},~\mathbb{T}^\mathrm{L0},~\mathbb{T}^\mathrm{T0}$ as
\bea
(\mathbb{T}^\mathrm{S0}_{lm} )_{\mu \nu} &=&\left(1-\frac{r^2}{L^2}\right)^{-1} (\mathbb{T}^\mathrm{L0}_{lm} )_{\mu \nu} - \frac{1}{2}r^2 (\mathbb{T}^\mathrm{T0}_{lm} )_{\mu \nu}, \label{change_1}
\eea
and
\bea
(\mathbb{T}^\mathrm{ur}_{lm} )_{\mu \nu} &=&(\mathbb{T}^\mathrm{Ru}_{lm} )_{\mu \nu} + 2 \left(1-\frac{r^2}{L^2}\right)^{-1}  (\mathbb{T}^\mathrm{L0}_{lm} )_{\mu \nu}. \label{change_2}
\eea

The electric parity de Sitter Teukolsky solution \eqref{electric_teukolsky} can now be  written as
\bea \label{electric_teukolsky_2}
h_{\mu \nu} &= &\sum_{m=-2}^{2} \left[ A_m^\mathrm{uu} (u,r) (\mathbb{T}^\mathrm{uu}_{2m} )_{\mu \nu}+ A_m^\mathrm{Ru} (\mathbb{T}^\mathrm{Ru}_{2m} )_{\mu \nu}  + A_m^{\mathrm{L0}} (\mathbb{T}^\mathrm{L0}_{2m} )_{\mu \nu}+ A_m^\mathrm{Eu}(u,r) (\mathbb{T}^\mathrm{Eu}_{2m} )_{\mu \nu} \right. \nonumber \\
&&\left. +A_m^\mathrm{T0} (u,r) (\mathbb{T}^\mathrm{T0}_{2m} )_{\mu \nu}+A_m^\mathrm{E1}(u,r) (\mathbb{T}^\mathrm{E1}_{2m} )_{\mu \nu}+ A_m^\mathrm{E2}(u,r) (\mathbb{T}^\mathrm{E2}_{2m} )_{\mu \nu} \right],
\eea
where via equations \eqref{change_1}--\eqref{change_2}, we have,
\begin{align}
A_m^\mathrm{Ru} &= A_m^{\mathrm{ur}},   \\
A_m^{\mathrm{L0}}&=2\left(1-\frac{r^2}{L^2}\right)^{-1}A_m^\mathrm{ur} + \left(1-\frac{r^2}{L^2}\right)^{-1} A_m^\mathrm{S0},\\
A_m^\mathrm{T0} &=-\frac{1}{2}r^2 A_m^\mathrm{S0},
\end{align}
The set of six functions  for each $m$, $A_m^\mathrm{uu},  A_m^\mathrm{Eu}, A_m^{\mathrm{ur}}, A_m^\mathrm{S0}, A_m^\mathrm{E1}, A_m^\mathrm{E2}$  are given above \eqref{Amuu}--\eqref{AmE2}. 

Under the diffeomorphism \eqref{diffeo} the choice
\bea \label{elec_diffeo_1}
\xi_u (x) &=& \sum_{l=0}^{\infty} \sum_{m=-l}^{l} \xi^{(u)}_{lm} (u,r) Y_{lm}(x^A), \\
\xi_r (x) &=& \sum_{l=0}^{\infty} \sum_{m=-l}^{l} \xi^{\mathrm{(R)}}_{lm} (u,r) Y_{lm}(x^A),  \\
\xi_A (x) &=& \sum_{l=1}^{\infty} \sum_{m=-l}^{l} \xi^{\mathrm{(E)}}_{lm} (u,r)  Y_A^{lm}(x^B),  \label{elec_diffeo_3}
\eea
generates electric parity terms alone.  Our aim is to exhibit that starting with the electric parity de Sitter Teukolsky waves \eqref{electric_teukolsky_2}, we get BBP electric parity waves in the form \eqref{eq:elecsol_BBP}.   Restricting ourselves to $l=2$ harmonics, a calculation shows that under \eqref{diffeo} the coefficients of various electric parity tensor harmonics transform as
\bea\label{tensor_harmonics_coefficient_transformation}
&& A_m^\mathrm{uu} \to  A_m^\mathrm{uu}-\frac{2\;r}{L^2}\;\xi^{\mathrm{(R)}}_{2m}+\frac{2\;r^3}{L^4}\;\xi^{\mathrm{(R)}}_{2m}+\frac{2\;r}{L^2}\;\xi^{(u)}_{2m}-2\partial_u\xi^{(u)}_{2m},\\&&
A_m^\mathrm{Ru}\to A_m^\mathrm{Ru}-\frac{2\;r}{L^2}\;\xi^{\mathrm{(R)}}_{2m}-\partial_r\xi^{(u)}_{2m}-\partial_u\xi^{\mathrm{(R)}}_{2m},\\&&
A_m^\mathrm{L0}\to A_m^\mathrm{L0}-2\;\partial_r\xi^{\mathrm{(R)}}_{2m},\\&&
\label{tensor_harmonics_coefficient_transformation_4}
A_m^\mathrm{Eu}\to A_m^\mathrm{Eu}-\xi^{(u)}_{2m}-\partial_u\xi^{\mathrm{(E)}}_{2m},\\&&
A_m^\mathrm{T0}\to A_m^\mathrm{T0}+6\;\xi^{\mathrm{(E)}}_{2m}+2\;r\;\xi^{(u)}_{2m}-2r \left(1- \frac{r^2}{L^2}\right)\xi^{\mathrm{(R)}}_{2m},\\&&
A_m^\mathrm{E1}\to A_m^\mathrm{E1}+\frac{2}{r}\;\xi^{\mathrm{(E)}}_{2m}-\xi^{\mathrm{(R)}}_{2m}-\partial_r\xi^{\mathrm{(E)}}_{2m},
\label{tensor_harmonics_coefficient_transformation_6}
\\& &
A_m^\mathrm{E2}\to A_m^\mathrm{E2}-\xi^{\mathrm{(E)}}_{2m}.
\label{tensor_harmonics_coefficient_transformation_7}
\eea

In the BBP electric parity solution \eqref{eq:elecsol_BBP} there are no 
$
\mathbb{T}^\mathrm{Ru}, \ \mathbb{T}^\mathrm{L0}, \ \mathbb{T}^\mathrm{T0}, \ \mathbb{T}^\mathrm{E1}
$ tensor harmonic terms. Thus, we need to set the coefficients of these four tensor harmonics to zero after the general diffeomorphism starting with the electric parity
de Sitter Teukolsky waves. Setting the coefficient of $\mathbb{T}^\mathrm{L0}$ to zero, we  
get a differential equation that determines $\xi^{\mathrm{(R)}}_{2m}$ upto an integration `constant' $c_1(u)$ as, 
\be
\xi^{\mathrm{(R)}}_{2m}=c_1-\frac{3 A_m}{8 r^4}-\frac{1}{2 r^3}\left(1+\frac{r^2}{L^2}\right) \dot{A}_m-\frac{\ddot{A}_m}{4 r^2}.
\ee
Next, upon setting the coefficient of $\mathbb{T}^\mathrm{E1}$ to zero, we
get another differential equation. This equation determines $\xi^{\mathrm{(E)}}_{2m}$ upto an integration `constant'  $c_2(u)$ as,
\be
\xi^{\mathrm{(E)}}_{2m}=r c_1+r^2 c_2+\frac{A_m}{8 r^3}+\frac{1}{8 r^2}\left(1-\frac{8 r^2}{3 L^2}\right) \dot{A}_m+\frac{1}{12 r}\left(1+\frac{2 r^2}{L^2}\right) \ddot{A}_m+\frac{1}{12} A_m^{(3)} r .
\ee
Next, upon setting the coefficient of $\mathbb{T}^\mathrm{Ru}$ to zero, we
get yet another differential equation. This equation determines $\xi^{(u)}_{2m}$
upto an integration `constant' $c_3(u)$ as,
\bea
\xi^{(u)}_{2m}=c_3-r \dot{c}_1-\frac{r^2 c_1}{L^2}+\frac{9}{8 L^2 r^2} A_m-\frac{1}{8 r^3}\left(1-\frac{16 r^2}{L^2}\right) \dot{A}_m-\frac{\ddot{A}_m}{4 r^2}-\frac{A_m^{(3)}}{4 r} .
\eea
Finally, upon setting the coefficient of $\mathbb{T}^\mathrm{T0}$ to zero, we get two algebraic relations involving different powers of $r$. One of these relations  determines  $c_3$ and the other determines $c_2$. We have,
\begin{align}
c_3&~=~-2 c_1,\\
c_2&~=~\frac{ \dot{A}_{m}}{4 L^4}+ \frac{1}{3} \dot{c}_1.
\label{relation_integral_constants}
\end{align}
In the end, we are left with only one function $c_1$ after setting the coefficients of $
\ \mathbb{T}^\mathrm{L0}, \ \mathbb{T}^\mathrm{E1}, \ \mathbb{T}^\mathrm{Ru}, \ \mathbb{T}^\mathrm{T0}$ tensor harmonics to zero. Choosing $c_1$ as
\be
c_1=-\frac{3 A_m}{2 L^4}+\frac{\ddot{A}_{m}}{8 L^2},
\ee
and identifying the combination $ -\frac{3 A_m}{4 L^2}+\frac{\ddot{A}_{m}}{12}$ as $a_m$, i.e., 
\be
a_m = -\frac{3 A_m}{4 L^2}+\frac{\ddot{A}_{m}}{12}, \label{iden_a_A}
\ee
the electric parity
de Sitter Teukolsky waves  gets  mapped to electric parity BBP waves. Note that in the presentation of the BBP electric parity waves only two derivatives of the functions $a_m$ appear whereas in the presentation of the electric parity de Sitter Teukolsky waves four-derivatives of the functions $A_m$ appear. Identification \eqref{iden_a_A} also explains this observation.

\section{Relation to Loganayagam-Shetye waves}\label{sec:ls_construction}
The positive cosmological constant has presented difficulties in thinking about the relation between gravity and quantum mechanics. The standard holographic picture inspired by the AdS/CFT correspondence that certain gravitational theories in the bulk of AdS are equivalent to certain non-gravitational quantum theories at the timelike boundary, does not seem to work straightforwardly. In asymptotically de Sitter spacetimes, the natural boundary is spacelike.  There are many ideas on de Sitter holography. For the purposes of the present paper,  linearised gravitational perturbations studied by Loganayagam and Shetye (LS) \cite{Loganayagam:2023pfb, Loganayagam:2023pfb_2} are of interest. A few words on the motivation behind their work are in order here.

The work of LS is motivated by the idea of  ``solipsistic'' holography \cite{Anninos:2011af,Nakayama:2011qh} for static patch observers.  In this approach, one considers an observer probing de Sitter spacetime. The worldline of such an observer is thought of as a timelike boundary on which one hopes to make precise a holographic dictionary. Such an observer influences and is influenced by its surroundings. In such a context, a gravitational observer can be thought of as an open quantum system constantly interacting with the rest of the spacetime. In an attempt to make such ideas precise,  LS consider
linearised gravitational waves in four-dimensional de Sitter spacetime building upon the earlier works \cite{Mukohyama:2000ui, Kodama:2003jz, Ishibashi:2004wx}. We are not concerned with the interpretation of their results or for that matter with the general idea of ``solipsistic'' holography, but rather with the linearised gravitational perturbations by themselves that they construct.  They work in the same set of coordinates as we work in, except that the ranges of the radial coordinate are different. It turns out that the  difference is only formal.  The boundary conditions for the linearised perturbations they consider are purely outgoing at the de Sitter horizon. These boundary conditions are naturally adapted to the set-ups we are interested in.

In their framework, de Sitter spacetime metric written in the outgoing Eddington-Finkelstein coordinates takes the form,
\begin{equation}
ds^2_{\mathrm{dS}}=- \left(1- \frac{r^2}{L^2}\right)du^2-2dudr+r^2\left(d\theta^2+\sin^2\theta d\phi^2\right).
\end{equation}
This form is the same as the Bondi form. The only difference is that Loganayagam and Shetye (LS) focus on the $r < L$ region whereas we are interested in the $r > L$ region. See Fig.~\ref{fig:dS}. However, the continuation from $r < L$ to $r > L$ is immediate as they do not necessarily require smoothness at the origin.  They only demand outgoing boundary conditions at the de Sitter horizon. 

 Outside the sources, using the master-variable framework of \cite{Mukohyama:2000ui, Kodama:2003jz, Ishibashi:2004wx}, they map the metric components for gravitational perturbations to two  potentials
  $\Phi_E(r,\w,l,m)$ and $\Phi_B(r,\w,l,m)$,
\begin{align}
& ds^2 =  ds^2_{\mathrm{dS}}+\sum_{l=0}^{\infty} \sum_{m=-l}^{l} \int \frac{d\w}{2\pi}e^{-i \frac{\w}{L} u } 
\left[\Psi_{uu}du^2+2\Psi_{ur}dudr+\Psi_{rr}dr^2+2r^2\Psi_s\gamma_{AB}\ dx^A dx^B\right] Y_{lm} \nonumber \\
&  \quad \quad +2\sum_{l=0}^{\infty} \sum_{m=-l}^{l} \int \frac{d\w}{2\pi}e^{-i \frac{\w}{L} u } \left(\frac{1}{L} \ du \ \Dp+dr\ \partial_r \right) \Phi_{B}\ (\vec r \times \vec \nabla)_A Y_{lm} dx^A\ , 
\label{pertubation_LS}
\end{align}
where 
\be
\Dp=\left(1-\frac{r^2}{L^2}\right)L\partial_r+i\w,
\ee 
and $\Psi$'s are written in terms of the electric potential $\Phi_E(r,\w,l,m)$ in the following.

Note that in this notation $\omega$ is dimensionless. The differential operator $(\vec r \times \vec \nabla)_A$ is a standard notation used, e.g., in electrodynamics \cite{Jackson:1998nia}. In terms of angular coordinates
\be
(\vec r \times \vec \nabla)_A = \left\{ -\frac{1}{\sin \theta} \partial_\phi, \partial_\theta \right\}.
\ee

The magnetic  potential $\Phi_B(r,\w,l,m)$ terms come together with the $(\vec r \times \vec \nabla)_A Y_{lm} $ vector harmonics in  \eqref{pertubation_LS}. The magnetic  potential $\Phi_B(r,\w,l,m)$  satisfies the following second order differential equation in $r$,
\begin{equation}
\frac{r^2}{L^2}\left(r^{2} D_{+}\left(r^{-2} D_{+} \Phi_B\right)+\omega^2 \Phi_B\right)-\left(1-\frac{r^2}{L^2}\right)(l-1)(l+2) \Phi_B=0.
\end{equation}

The terms that come together with the scalar spherical harmonics in equation \eqref{pertubation_LS} 
are written in terms of the electric   potential $\Phi_E(r,\w,l,m)$,
\begin{align}
&\Psi_{uu} =\Dp\Phi_E-\w^2 \frac{r}{L} \ \Phi_E+\frac{l(l+1)L}{2r} \left(1-\frac{r^2}{L^2}\right)\Phi_E,& \\
&\Psi_{ur} =\left(1-\frac{r^2}{L^2}\right)^{-1}\left\{-i\w \frac{r}{L}\left(\Dp\Phi_E+\frac{L}{r}\Phi_E\right)+\Psi_{uu}\right\} ,&\\
&\Psi_{rr}=2\left(1-\frac{r^2}{L^2}\right)^{-1}\Psi_{ur},&\\
&\Psi_s=\frac{1}{2}\Dp\Phi_E+\frac{l(l+1)L}{4r}\Phi_E,&
\end{align}
where $\Phi_E$ satisfies the following second order differential equation in $r$,
\begin{align} \label{Phi_E_eq}
\frac{r^2}{L^2}\left( D_{+}\left( D_{+} \Phi_E\right)+\omega^2 \Phi_E\right)-\left(1-\frac{r^2}{L^2}\right)l(l+1) \Phi_E=0.
\end{align}
Using \eqref{Phi_E_eq} expressions for $\Psi_{rr}$ and $\Psi_{ur}$ can be simplified,
\begin{align}
&\Psi_{rr}= \frac{L}{r}\partial_r\left(r^2\partial_r\Phi_E\right), &\\
&\Psi_{ur} =  \frac{L}{2r} \left(1-\frac{r^2}{L^2}\right) \partial_r\left(r^2\partial_r\Phi_E\right).& 
\end{align}

Linearised Einstein's equations with outgoing boundary conditions are satisfied for equation \eqref{pertubation_LS} provided,
\begin{align}
\Phi_E(r,\w,l,m) &=G_{E}(r,\w,l) \ \tilde{e}_{l m}(\w), \\
\Phi_B(r,\w,l,m) &=G_{B}(r,\w,l) \ \tilde{d}_{l m}(\w),
\end{align}
where $\tilde e_{l m}(\w), \tilde d_{l m}(\w)$ are arbitrary functions of $\omega$ and functions $G_E$ and $G_B$ are given as,
\be
G_E = \left(\frac{r}{L}\right)^{l+1}\left(1+ \frac{r}{L}\right)^{-i\w}\frac{\Gamma\left(\frac{l+2-i\w}{2}\right)\Gamma\left(\frac{l+1-i\w}{2}\right)}{\Gamma\left(1-i\w\right)\Gamma\left(l+\frac{1}{2}\right)}{}_2F_1\left(\frac{l+1-i\w}{2}, \frac{l+2-i\w}{2};1-i\w;1-\frac{r^2}{L^2}\right), \label{G_E}
\ee
\be
G_B =  \left(\frac{r}{L}\right)^{l+2}\left(1+ \frac{r}{L}\right)^{-i\w}\frac{\Gamma\left(\frac{l+1-i\w}{2}\right)\Gamma\left(\frac{l+2-i\w}{2}\right)}{\Gamma\left(1-i\w\right)\Gamma\left(l+\frac{1}{2}\right)}{}_2F_1\left(\frac{l+1-i\w}{2}, \frac{l+2-i\w}{2};1-i\w;1-\frac{r^2}{L^2}\right), \label{G_B}
\ee
where ${}_2F_1(a,b;c;z)$ is the Gauss hypergeometric function. Functions $\tilde e_{l m}(\w),\tilde d_{l m}(\w)$  are related to the (time-dependent) multipole moments for the potentials $\Phi_E$ and $\Phi_B$.

Note that the hypergeometric functions appearing in \eqref{G_E} and \eqref{G_B} are the same. Moreover, note that the arguments $a,b,c$ in \eqref{G_E} for the hypergeometric function ${}_2F_1(a,b;c;z)$ are such that 
\be
a -b = \frac{1}{2}.
\ee	
In such cases there exist ``quadratic transformation formulas'' \cite{abramowitz1948handbook}. Using formula 15.3.32 (page 561) of \cite{abramowitz1948handbook} we have,
\begin{align} \label{HG_identity}
{}_2F_1\left(a,1-a;c;z\right)=(1-z)^{c-1}(1-2z)^{a-c}{}_2F_1\left(\tfrac{1}{2}c-\tfrac{1}{2}a,\tfrac{1}{2}c- \tfrac{1}{2}a+\tfrac{1}{2};c;\frac{4z^2-4z}{(1-2z)^2}\right).
\end{align}
With the identification
\begin{align}
a&=-l, & c&=1-i \omega, & z&=\frac{r-L}{2r},
\end{align} 
the arguments for ${}_2F_1$ function on the right hand side of identity \eqref{HG_identity} match with the arguments for ${}_2F_1$ function appearing in \eqref{G_E} and \eqref{G_B}. The key point to note is that via identity \eqref{HG_identity}, the first argument of ${}_2F_1$ function on the left hand side is a negative integer. As a result, the hypergeometric series for ${}_2F_1$ function on  the left hand side of identity \eqref{HG_identity} reduces to a polynomial of degree $l$. 

Since we are only interested in the quadrupolar perturbations, we set $l=2$ in the above expressions. Upon using the ``duplication formula'' for the Gamma function $\Gamma(z)$ \cite[eq.~6.1.18, page 256]{abramowitz1948handbook}
\be
\Gamma(2z) = (2 \pi)^{-\frac{1}{2}} 2^{2z - \frac{1}{2}} \Gamma(z) \Gamma\left(z + \frac{1}{2} \right),
\ee
with $z = \frac{3}{2} - \frac{i}{2} \omega$, functions $G_E$ and $G_B$ simplify to
\begin{align}
G_E(r,\w,2) &= -\frac{1}{3}(1+\w^2)-\frac{i\w L}{r}+\frac{L^2}{r^2} ,\\
G_B(r,\w,2) &= \frac{r}{L} \left(-\frac{1}{3}(1+\w^2)-\frac{i\w L}{r}+\frac{L^2}{r^2} \right).
\end{align}
It is quite remarkable that these functions reduce to polynomials in $r$. Next, we  define the functions $d_{lm}(u)$ and $e_{lm}(u)$ as the Fourier transforms of $\tilde d_{lm}(\w)$ and $\tilde e_{lm}(\w)$, respectively, 
\begin{align}
d_{lm}(u)&=\int_{-\infty}^{\infty} \frac{d\w}{2\pi} \ e^{-i \frac{\w}{L} u } L \ \tilde d_{lm}(\w), &
e_{lm}(u)&=\int_{-\infty}^{\infty} \frac{d\w}{2\pi} \ e^{-i \frac{\w}{L} u }L^{3} \ \tilde e_{lm}(\w). 
\end{align}
The functions $\{\tilde d_{lm}(\w), \tilde e_{lm}(\w)\}$ and $\{d_{lm}(u), e_{lm}(u)\}$ do not have the same dimensions. Different powers of $L$ are inserted in going from $\{\tilde d_{lm}(\w), \tilde e_{lm}(\w)\}$  to $\{d_{lm}(u), e_{lm}(u)\}$ so that we can straightforwardly take the $L \to \infty$ limit in the final solutions.  To further simplify the notation, $d_{lm}(u)$ and $e_{lm}(u)$ functions with $l=2$, namely $d_{2m}(u)$ and $e_{2m}(u)$,  are simply denoted as $d_m(u)$ and $e_{m}(u)$. 

With these simplifications, the $l=2$ magnetic  parity solution obtained from \eqref{pertubation_LS} can be written as,
\begin{align}
\label{pertubation_LS_magnetic}
h_{\mu \nu}^{\text{(B)}} &=  \sum_{m=-2}^{2} 
\left[\frac{1}{r^2}\left(1-\frac{2r^2}{3 L^2}-\frac{r^4}{3 L^4}\right) d_m+\frac{1}{r}\left(1-\frac{r^2}{3 L^2}\right) \dot{d}_{m}+\frac{2}{3}\left(1+\frac{r^2}{2 L^2}\right) \ddot{d}_{m}+\frac{r\; d^{(3)}_m}{3}\right] (\mathbb{T}^{\mathrm{Bu}}_{2m})_{\mu \nu} \nonumber \\ 
&   + \sum_{m=-2}^{2}\left[ \frac{1}{r^2}\left(1+\frac{r^2}{3 L^2}\right) d_m-\frac{\ddot{d}_{m}}{3}\right](\mathbb{T}^{\mathrm{B1}}_{2m})_{\mu \nu},
\end{align}
where we have used the magnetic parity tensor harmonics introduced in section \ref{sec:dS_Teukolsky}. As before, the first and the second $u$ derivatives of the functions $d_m(u)$ are denoted as $\dot{d}_m$ and $\ddot{d}_m$ and the higher derivatives are denoted with superscripts, e.g., $d^{(3)}_m$. The $l=2$ electric  parity solution obtained from \eqref{pertubation_LS} can similarly be written as,
\begin{align}
\label{pertubation_LS_electric}
h_{\mu \nu}^{\text{(E)}} &=  \sum_{m=-2}^{2}\left[\frac{1}{r^3}\left(1-\frac{r^2}{L^2}\right)^2 e_m+\frac{1}{r^2}\left(1-\frac{5 r^2}{3 L^2}\right)\dot{e}_m+\frac{1}{r}\left(1-\frac{4 r^2}{3 L^2}\right)\ddot{e}_m+\frac{2}{3} e_m^{(3)}+\frac{1}{3} r e_m^{(4)} \right](\mathbb{T}^\mathrm{uu}_{2m} )_{\mu \nu}\nonumber\\
& + \sum_{m=-2}^{2}	\left[ \frac{1}{r}\left(1+\frac{r^2}{L^2}\right) e_m+\left(1+\frac{4 r^2}{3 L^2}\right) \dot{e}_m-\frac{1}{3} r^2 e_m^{(3)} \right] (\mathbb{T}^\mathrm{T0}_{2m} )_{\mu \nu}+\sum_{m=-2}^{2}  \frac{2 e_m}{r^3} (\mathbb{T}^\mathrm{L0}_{2m} )_{\mu \nu} \nonumber \\ &+ \sum_{m=-2}^{2} \left[\frac{1}{r^3}\left(1-\frac{r^2}{L^2}\right) e_m\right]  (\mathbb{T}^\mathrm{Ru}_{2m} )_{\mu \nu}.
\end{align}

Both the electric and  magnetic parity solutions are written in a gauge similar to the Regge-Wheeler gauge, where the terms involving the $(\mathbb{T}^\mathrm{E2}_{2m} )_{\mu \nu}$ and $(\mathbb{T}^\mathrm{B2}_{2m} )_{\mu \nu}$ tensor harmonics do not appear. To compare with de Sitter Teukolsky waves from sections \ref{sec:magnetic_Teukolsky} and \ref{sec:electric_Teukolsky}, we need to perform a gauge transformation as before. 

\paragraph{Magnetic Parity.}

Starting with the magnetic parity de Sitter Teukolsky waves \eqref{magnetic_teukolsky} 
we can get rid of the coefficient of $ (\mathbb{T}^{\mathrm{B2}}_{2m})_{\mu \nu}$ by appropriately choosing $\xi^{\mathrm{(B)}}_{2m} (u,r)$ via \eqref{transform_magnetic_1},
\be
\xi^{\mathrm{(B)}}_{2m}  = \frac{3}{r^2}\left(1 + \frac{2r^2}{3L^2} + \frac{r^4}{L^4}  \right) B_m+ \frac{3}{r} \left(1+ \frac{r^2}{L^2}\right) \dot{B}_{m}+ 2\left( 1 + \frac{r^2}{2L^2} \right) \ddot{B}_{m}+  r B_m^{(3)}  . 
\ee
Furthermore, identifying,
\be
B_m(u) = \frac{1}{3}\int^u d_m(u') du' ,
\ee
we get  the magnetic parity solution in the form of equation \eqref{pertubation_LS_magnetic}.

\paragraph{Electric parity.}
Once again, for the electric parity case, the discussion is almost identical, but a little more involved. 
Starting with the electric parity de Sitter Teukolsky waves \eqref{electric_teukolsky} 
we can get rid of the coefficient of $ (\mathbb{T}^{\mathrm{E2}}_{2m})_{\mu \nu}$ by appropriately choosing $\xi^{\mathrm{(E)}}_{2m} (u,r)$ via equation \eqref{tensor_harmonics_coefficient_transformation_7},
\bea 
\xi^{\mathrm{(E)}}_{2m}&=&\frac{1}{8 r^3}\left(1-\frac{3 r^2}{L^2}\right) A_m+\frac{1}{8 r^2}\left(1-\frac{8 r^2}{3 L^2}-\frac{2 r^4}{ L^4}\right) \dot{A}_m+\frac{1}{8 r}\left(1-\frac{2 r^2}{L^2}\right) \ddot{A}_m\nonumber\\
&&+\frac{1}{12}\left(1+\frac{r^2}{2 L^2}\right) A_m^{(3)}+\frac{1}{24} r A_m^{(4)} .
\eea
Next, we can get rid of the coefficient of $ (\mathbb{T}^{\mathrm{E1}}_{2m})_{\mu \nu}$ by appropriately choosing $\xi^{\mathrm{(R)}}_{2m} (u,r)$ via equation \eqref{tensor_harmonics_coefficient_transformation_6},
\begin{align}
\xi^{\mathrm{(R)}}_{2m}=&-\frac{3}{8 r^4}\left(1+\frac{3 r^2}{L^2}\right) A_m-\frac{1}{2 r^3}\left(1-\frac{r^2}{L^2}\right) \dot{A}_m-\frac{1}{8 r^2}\left(1+\frac{10 r^2}{3 L^2}\right) \ddot{A}_m+\frac{1}{24} A_m^{(4)}.
\end{align}
Then, we can get rid of the coefficient of $ (\mathbb{T}^{\mathrm{Eu}}_{2m})_{\mu \nu}$ by appropriately choosing $\xi^{(u)}_{2m} (u,r)$ via equation~\eqref{tensor_harmonics_coefficient_transformation_4}, 
\begin{align}
\xi^{(u)}_{2m}=& -\frac{1}{8 r^3}\left(1-\frac{7 r^2}{L^2}\right) \dot{A}_m-\frac{1}{8 r^2}\left(1-\frac{20 r^2}{3 L^2}-\frac{10 r^4}{3 L^4}\right) \ddot{A}_m-\frac{1}{8 r}\left(1-\frac{10 r^2}{3 L^2}\right) A_m^{(3)}\nonumber\\
&-\frac{1}{12}\left(1+\frac{r^2}{2 L^2}\right) A_m^{(4)}-\frac{1}{24} r A_m^{(5)}.
\end{align}
Finally, identification of the combination $-\frac{1}{4}\left(\frac{9}{L^2} A_m- \ddot{A}_{m}\right)$ as $e_m$, i.e., 
\bea
e_m=-\frac{1}{4}\left(\frac{9}{L^2} A_m- \ddot{A}_{m}\right),
\eea
maps the electric parity
de Sitter Teukolsky waves to electric parity waves in the form \eqref{pertubation_LS_electric}.

\section{Relation to Comp\`ere-Hoque-Kutluk waves}
\label{sec:CHK_dS}

Expanding on the work of Ashtekar, Bonga, and Kesavan (ABK) \cite{deVega:1998ia, Ashtekar:2015lla,  Ashtekar:2015ooa, Ashtekar:2015lxa}, Comp\`ere, Hoque, and Kutluk (CHK) 
\cite{Compere:2023ktn} obtained  expressions for the metric perturbations in Bondi gauge around de Sitter spacetime generated by localised sources. They also made more precise the meaning of quadrupolar truncation in the multipolar expansion. The aim of this section is to show that in the purely quadrupolar truncation, the CHK solutions are equivalent to the de Sitter Teukolsky solutions of section \ref{sec:dS_Teukolsky}.

To explain CHK solutions in Bondi gauge, we need to first consider linearised gravity in the future Poincar\'e patch. We need to introduce a significant notation before we can write the CHK solutions. In Poincar\'e patch, the background de Sitter metric is
\be
ds^2 = a(\eta)^2( - d \eta^2 + dx_i dx_i), 
\ee
where
\be
a(\eta) = - \frac{1}{H \eta}, 
\ee
with $H = \frac{1}{L} = \sqrt{\frac{\Lambda}{3}}.$ The coordinates $x_i$ are in the range $(- \infty, \infty)$ and the coordinate $\eta$ is in the range $(-\infty, 0)$, with $\eta = 0$ at the future infinity and $\eta \to -\infty$ at the cosmological horizon (the boundary of the Poincar\'e patch).

In refs.~\cite{deVega:1998ia, Ashtekar:2015lla,  Ashtekar:2015ooa, Ashtekar:2015lxa}, ABK  considered solving inhomogeneous equations in the Poincar\'e patch. In order to discuss the solutions, we need to briefly discuss definitions and some properties of the moments of the \textit{source} stress tensor. The following definitions for source moments turn out to be convenient to work with \cite{Ashtekar:2015lxa, Date:2015kma, Compere:2023ktn}. We introduce moments of the energy density,
\begin{align} 
Q^{(\rho)}(\eta)&= \int d^3x\;a(\eta) T_{00}, \\ 
Q_i^{(\rho)}(\eta)&= \int d^3x\;a^{2}(\eta) T_{00} x_i, \\
Q_{ij}^{(\rho)}(\eta)&= \int d^3x\; a^{3}(\eta) T_{00} x_i x_j,
\label{rho-moments}
\end{align} 
of the momentum density,
\begin{align} 
& P_{i}(\eta)= \int d^3x\;a(\eta) T_{0i}, \\
& P_{i \vert j}(\eta)= \int d^3x\;a^{2}(\eta) T_{0i} x_j \\
& P_{i \vert jk}(\eta) = \int d^3x\;a^{3}(\eta) T_{0i} x_jx_k,
\end{align} 
of the spatial part of the stress tensor,
\begin{align} 
&S_{ij}(\eta)= \int d^3x\;a(\eta)  T_{ij}, \\
&S_{ij \vert k}(\eta)= \int d^3x\;a^{2}(\eta)  T_{ij}  x_k,
\end{align} 
and of the spatial trace of the  stress tensor,
\begin{align} 
&Q^{(p)}(\eta)= \int d^3x\;a(\eta)  \delta^{kl}T_{kl}, \\
&Q^{(p)}_{i}(\eta)= \int d^3x\;a^{2}(\eta)  \delta^{kl}T_{kl}  x_i, \\
&Q^{(p)}_{ij}(\eta)= \int d^3x\;a^{3}(\eta)  \delta^{kl}T_{kl}  x_ix_j.
\label{p-moments}
\end{align} 
In these equations, $T_{00}(\eta, x_i)$ and $T_{0i}(\eta, x_i)$ are the $\eta\eta$ and the $\eta i$ components the stress tensor respectively. Raising and lowering of $i$, $j$ indices are done with the three-dimensional  flat cartesian metric $\delta_{ij}$. We define  angular momentum $J_i$ as the odd parity dipole moment of the momentum density
\be
J_i = \epsilon_{ijk} P_{j|k}. \label{def_Ji}
\ee
We define symmetric and trace-free odd parity quadrupole moments $J_{ij}$ and $K_{ij}$ as
\be \label{JandK}
J_{ij} = \frac{4}{3} P_{k \vert l(i}\epsilon_{j)kl}, \qquad K_{ij} = \frac{4}{3} \epsilon_{kl(i}S_{j)k \vert l}.
\ee

The conservation of the source stress tensor relate these moments. For non-zero cosmological constant $\Lambda$, coordinates $\{ \eta, x_i \}$ give rise to a good differential structure, but in order to consider $\Lambda \to 0$ limit, we need to switch to another differential structure. A convenient choice is proper time $t$ and $x_i$. At fixed $x_i$,  
\be
dt^2  = a(\eta)^2 d \eta^2 \implies \eta = - \frac{1}{H} e^{-Ht}. 
\ee
It follows that $\partial_\eta = e^{Ht} \partial_t$. 
The resulting equations  from the conservation of the stress tensor relating various moments introduced above in the $\{t, x_i\}$ coordinates are carefully explained in refs.~\cite{Ashtekar:2015lxa, Date:2015kma, Compere:2023ktn}; we do not repeat them here.  Instead, equations relating various moments in the $\{u, x_i\}$ coordinates are most useful to us.

The coordinate transformation from Poincar\'e patch coordinates $\{ \eta, x_i \}$ to Bondi coordinates $\{ u,r,\theta, \phi \}$ is 
\begin{equation} 
u=-\frac{1}{H} \ln \bigg(H(\rho - \eta)\bigg), 
~~~~ r=-\frac{\rho}{H\eta}, 
\end{equation} 
where  $\rho := \sqrt{x_i x_i}$. 
Linearised fields are generated by sources at the retarded time $t_{\mbox{\small{ret}}}$. The retarded time is related to the Poincar\'e patch coordinates as, $\eta_{\mbox{\small{ret}}}:= \eta -\rho= -\frac{1}{H} e^{- Ht_{\mbox{\small{ret}}}}$.
It then follows that $\partial_{t_{\mbox{\small{ret}}}}=\partial_u$. As a result, in terms of the $u$ derivatives the equations relating various moments of the stress tensor take the same form as they take in terms of the $t$ derivatives. Therefore, we have \cite{Compere:2023ktn},
\begin{align} \label{iden-1}
&\partial_u Q^{(\rho)} = - H Q^{(p)}, & 
&Q^{(p)}= \delta^{ij} S_{ij},\\
\label{iden-2}
&\partial_{u} P_{i}=-HP_{i}, &
& \partial_u Q_i^{(\rho)} = - P_i + H Q_i^{(\rho)}  - HQ_i^{(p)}, \\
\label{iden-3}
&\partial_u J_i=\epsilon_{ijk}\partial_u P_{j|k} = 0,&
&\partial_u P_{i|j}=-S_{ij},  \\
\label{iden-4}
&S_{ij|j}=-\frac{1}{2}S_{jj|i}=-\frac{1}{2}Q_{i}^{(p)}, &
&\partial_{u} P_{i|jk}= H P_{i|jk} -S_{ij\vert k}-S_{ik \vert j},\\
&2P_{(i|j)}=-\partial_{u}Q_{ij}^{(\rho)}+2HQ_{ij}^{(\rho)}-HQ_{ij}^{(p)}, &
&S_{ij}=\frac{1}{2}\partial_{u}(\partial_{u}Q_{ij}^{(\rho)}-2H Q_{ij}^{(\rho)}+H Q_{ij}^{(p)}).
\label{iden-5}\end{align}

We are now in a position to write the CHK solution in Bondi gauge. Following CHK, we write the background plus the linear perturbation on it: $g_{\mu \nu} = \bar g_{\mu \nu}+H_{\mu \nu}$. Components $g_{AB} $ are,
\begin{equation}
g_{AB} = \bar g_{AB}+ H_{AB} = r^2 q_{AB}+ r C_{AB} + \frac{1}{r} E_{AB} + \mathcal{O}\left(\frac{1}{r^2}\right),
\end{equation}
where $q_{AB}$, $C_{AB}$ and $E_{AB}$ tensors are 
\begin{align}
q_{AB} &= \mathring{q}_{AB}
+ e^i_{\langle A} e^j_{B \rangle} \bigg(\partial_u \zeta_{ij}+2H^2 \partial_u Q^{(\rho+p)}_{ij}
+2H^2 \epsilon_{i kl}n_k(K_{jl}+H\int^u du'K_{jl}(u'))\bigg), \label{qABBondi}\\
C_{AB} &= 
e_{\langle A}^i e_{B \rangle}^j \left( 3\zeta_{ij} +2 (\partial_u^2 -H^2)Q^{(\rho+p)}_{ij}+2\epsilon_{ikl}n_k (\partial_u+H)K_{lj}\right),\\
E_{AB} & = 2 e_{\langle A}^i e_{B \rangle}^j \left( Q^{(\rho+p)}_{ij} + \epsilon_{ikl}n_k J_{lj}\right), 
\end{align}
and where $\mathring{q}_{AB}$ is the round metric on the unit two-sphere and $n^i $ is the unit vector in the radial direction $n^i= \left\{ \sin \theta \cos \phi, \sin \theta \sin \phi, \cos \theta \right\}.$
The angular bracket  expression $e_{\langle A}^i e_{B \rangle}^j$ stands for,
\be
e_{\langle A}^i e_{B \rangle}^j = \frac{1}{2} \left( e_{A}^i e_{B}^j +  e_{B}^i e_{A}^j -  \mathring{q}_{AB} (\delta^{ij}- n^{i}n^{j}) \right),
\ee
where $e^i_{A} = \partial_A n^i$. 
Moreover, $\zeta_{ij}$ is related to $Q^{(\rho+p)}_{ij}$ via the relation
\be
\partial_u^2\zeta_{ij} - 3 H^2 \zeta_{ij} =  - 2 H^4 Q^{(\rho+p)}_{ij},\label{zeta-eq}
\ee
where $Q^{(\rho+p)}_{ij}$ are moments of the combination $T_{00} + \delta^{ij} T_{ij}$, cf.~\eqref{rho-moments}--\eqref{p-moments}.

The component $g_{uu} = \bar g_{uu}+H_{uu}$ reads
\begin{align}
g_{uu} &=-1+H^2 r^2+(\delta^{ij}-3n^i n^j)\partial_u (\zeta_{ij}+2H^2 Q^{(\rho+p)}_{ij}) 
+\frac{2M}{r}+\frac{2N}{r^2}+\frac{(3n^{i}n^{j}-\delta^{ij})Q^{(\rho+p)}_{ij}}{r^3},
\end{align}
where 
\begin{align}
M = &~ Q^{(\rho)}-2 Q^{(p)} -3n^{i}(P_{i}-HQ_{i}^{(\rho)}-H^2 P_{i \vert kk})+3 n^{i}n^{j}(2 S_{ij}-H P_{i|j})\nonumber\\
&~ + (\delta^{ij}-3n^{i} n^{j}) (3 H P_{i \vert j}-3H^2 Q^{(\rho)}_{ij} +3 H^2 Q^{(p)}_{ij}+H \partial_u Q^{(p)}_{ij}-\partial_u^2 Q^{(p)}_{ij}),\\
N = &~n_i (Q^{(\rho)}_i+H P_{i \vert kk})-(\delta_{ij}-3 n_{i}n_{j}) \partial_{u} Q_{ij}^{(\rho +p)}.
\end{align}
Finally, $H_{uA}$ components are given as 
\begin{align}
H_{uA} = &~ -e_A^i n^j (H^{-2}\partial_u^2 \zeta_{ij}+2\partial_u^2 Q^{(\rho+p)}_{ij}+2 \epsilon_{i kl}n_k(\partial_u+H) K_{jl}) 
\nonumber\\
&~+ \frac{2 e_A^i N_i }{r}+\frac{ 3e_A^i n^j ( Q_{ij}^{(\rho+p)}+ \epsilon_{ikl}n_k J_{lj})}{r^2}, 
\end{align}
where,
\begin{align}
N_i &= Q_i^{(\rho)}+H P_{i \vert kk} +n^j (\epsilon_{ijk}J_k+2\partial_u Q^{(\rho+p)}_{ij})-2 \epsilon_{ijk}n_j n_l (K_{kl}- H J_{kl}).
\end{align}
Compared to equations (3.45)--(3.53) of \cite{Compere:2023ktn}, we have set   $\mathring\xi^u$ and $\mathring\xi^A$ terms to zero. These terms represent the $\Lambda$-BMS asymptotic symmetries, which obey the equations, 
\be
\partial_u \mathring{\xi}^u-\frac{1}{2}\mathring{D}_A \mathring{\xi}^A =0,\qquad \partial_u \mathring{\xi}^A+H^2 \mathring{q}^{AB}\partial_B \mathring{\xi}^u=0.
\ee
It is consistent to set $\mathring\xi^u$ and $\mathring\xi^A$ to zero. Solutions with zero $\mathring\xi^u$ and $\mathring\xi^A$ versus non-zero $\mathring\xi^u$ and $\mathring\xi^A$ are related by 
$\Lambda$-BMS asymptotic symmetries. 

Except for the expression for the mass aspect function $M(u, x^A)$, all other terms cleanly separate into monopole, dipole, and quadrupole terms. We can achieve the same for the mass aspect  function $M(u, x^A)$, using identities \eqref{iden-1}--\eqref{iden-5}.  We can write,
\be 
M= Q^{(\rho)}-H\delta^{ij} P_{i|j}-3n^{i}(P_{i}-HQ_{i}^{(\rho)}-H^2 P_{i \vert kk})+(3n^i n ^j-\delta^{ij})(\partial_{u}^{2}Q_{ij}^{(\rho +p)}-H^{2}Q_{ij}^{(\rho +p)}). \label{simplifiedM}
\ee

The solution is very cumbersome. Let us separately analyse the monopole, dipole and quadrupole terms.

\subsection{Monopole and dipole truncations}
The monopole truncation $H_{\mu \nu}\big{|}_{l=0}$ of the CHK solution is simply,
\be
H_{uu}\big{|}_{l=0}=\frac{2}{r}\left(Q^{(\rho)}-H\delta^{ij} P_{i|j}\right),
\ee
with all other components of $H_{\mu \nu}\big{|}_{l=0}$ zero. The mass aspect is independent of $u$, i.e.,  
\be
\partial_u\left(Q^{(\rho)}-H\delta^{ij} P_{i|j}\right)=0,
\ee
which follows from identities \eqref{iden-1} and \eqref{iden-3}. This truncation of $H_{\mu \nu}$ satisfies linearised Einstein's equation. 

The magnetic parity dipole truncation  $H_{\mu \nu}^{\text{(B)}}\big{|}_{l=1}$ of the CHK solution is,
\be
H_{uA}^{\text{(B)}}\big{|}_{l=1}= \frac{2 e_A^i n^j \epsilon_{ijk}}{r}J_k, 
\ee
where $J_i=\epsilon_{ijk}P_{j|k}$, cf.~\eqref{def_Ji}, and all other components of $H_{\mu \nu}^{\text{(B)}}\big{|}_{l=1}$ are zero. In terms of the magnetic parity tensor harmonics, the perturbation is, 
\bea
H_{\mu \nu}^{\text{(B)}}\big{|}_{l=1}=\sum_{m=-1}^{1}\frac{2j_m}{r} \left(\mathbb{T}^{\text{Bu}}_{1m} \right)_{\mu\nu},
\eea
with the identification
\begin{align}
j_1 &= \sqrt{\frac{2 \pi}{3}} (J_1 - i J_2) , & j_{-1} &= - \sqrt{\frac{2 \pi}{3}} ( J_1 + i J_2) , & j_0 &= -2 \sqrt{\frac{ \pi}{3}} J_3. 
\end{align}
From equation \eqref{iden-3} we note that $\partial_u J_i= 0$.  This perturbation satisfies linearised Einstein's equation.

The electric parity dipole truncation  $H_{\mu \nu}^{\text{(E)}}\big{|}_{l=1}$ of the CHK solution  has $H_{uu}^{\text{(E)}}\big{|}_{l=1}$ and $H_{uA}^{\text{(E)}}\big{|}_{l=1}$ components non-zero,
\bea \label{electric_dipole_H_1}
H_{uu}^{\text{(E)}}\big{|}_{l=1}&=& -\frac{6}{r}(P_{i}-HQ_{i}^{(\rho)}-H^2 P_{i \vert kk})n^{i}+\frac{2}{r^2}( Q^{(\rho)}_i+H P_{i \vert kk})n^{i}, \\
H_{uA}^{\text{(E)}}\big{|}_{l=1}&=&\frac{2 e_A^i }{r}(Q_i^{(\rho)}+H P_{i \vert kk} ),
\eea 
and all other components of $H_{\mu \nu}^{\text{(E)}}\big{|}_{l=1}$ are zero. This perturbation can be written more simply by introducing
\be
v_i=Q_i^{(\rho)}+H P_{i \vert kk}.
\ee  We note from identities \eqref{iden-2} and \eqref{iden-4} that,
\bea
\partial_u v_i &=& \partial_u Q_i^{(\rho)}+ H \partial_u P_{i \vert kk} \\
&=& - P_i + H Q_i^{(\rho)} -H Q_i^{(p)} +  H^2 P_{i|kk} - H S_{ik\vert k} - H S_{ik \vert k} \\
&=& - P_i + H Q_i^{(\rho)}+  H^2 P_{i|kk}. 
\eea
Using this identity in equation \eqref{electric_dipole_H_1}, we see that the perturbation simplifies as,
\bea
H_{uu}^{\text{(E)}}\big{|}_{l=1}&=& \frac{6}{r} (\partial_u v_i) n^{i}+\frac{2}{r^2} v_i n^{i}, \\
H_{uA}^{\text{(E)}}\big{|}_{l=1}&=&\frac{2 e_A^i }{r}v_i, 
\eea
In terms of the electric parity tensor harmonics, it takes the form, 
\be
H_{\mu\nu}^{\text{(E)}}\big{|}_{l=1}=\sum_{m=-1}^{1} \left(\frac{2 k_m}{r^2}+\frac{6\dot{k}_m}{r} \right)\left( \mathbb{T}^{\text{uu}}_{1m} \right)_{\mu\nu}+ \sum_{m=-1}^{1} \frac{2 k_m}{r} \left(\mathbb{T}^{\text{Eu}}_{1m} \right)_{\mu\nu},
\ee 
where
\begin{align}
k_1 &= -\sqrt{\frac{2 \pi}{3}} (v_1 - i v_2) , & k_{-1} &=  \sqrt{\frac{2 \pi}{3}} ( v_1 + i v_2) , & k_0 &= 2 \sqrt{\frac{ \pi}{3}} v_3.\label{l=1_elec_relations}
\end{align}
This perturbation satisfies linearised Einstein's equations provided $-k_m+L^2\ddot{k}_m=0$. Via \eqref{l=1_elec_relations}, this relation is equivalent to $-v_i+L^2\ddot{v}_i=0$, which follows from  \eqref{iden-1}, \eqref{iden-2} and \eqref{iden-4}.

\subsection{Magnetic parity quadrupole truncation}
Having identified the monopole and the dipole parts, in this section we analyse the  purely quadrupolar truncation of the CHK solution. The aim is to illustrate that starting from the de Sitter Teukolsky waves we can get the CHK wave solutions. Let us start with the magnetic parity waves. 

The magnetic parity CHK quadrupole solution $H_{\mu \nu}^{\text{(B)}}\big{|}_{l=2} $is,
\begin{align}
H_{uu}^{\text{(B)}}\big{|}_{l=2} &= 0,\\
H_{uA}^{\text{(B)}}\big{|}_{l=2}  &=e_A^i n^j n_k \epsilon_{i kl}\left(- 2(\partial_u+H) K_{jl}- \frac{4}{r}\left(  K_{jl}- H J_{jl}\right)+\frac{3}{r^2}J_{lj}\right), \\
H_{AB}^{\text{(B)}}\big{|}_{l=2}  &= e^i_{\langle A} e^j_{B \rangle}n_k \epsilon_{ikl}\left(2r^2 H^2 \left(K_{jl}+H\int^u du'K_{jl}(u')\right)+ 2r (\partial_u+H)K_{lj} + \frac{2}{r}   J_{lj}\right), 
\end{align}
where $J_{ij}$ and $K_{ij}$ are the symmetric and trace-free magnetic parity quadrupole moments introduced in \eqref{JandK}. They are related as
\be\label{JandK_relation}
(\partial_u - H)J_{ij} = - K_{ij}.
\ee
Substituting for $K_{ij}$ we get a form of the magnetic parity CHK quadrupole solution written only in terms of $J_{ij}(u)$. 

In order to write this perturbation in terms of the magnetic parity tensor harmonics, we start by considering the combinations  $e_A^i n^j  \epsilon_{ikl}n_k J_{lj}$ and $e^{i}_{\langle A}e^{j}_{ B \rangle} \epsilon_{ikl}n_k J_{lj}$. A calculation shows that,
\begin{align}
2 e_A^i n^j\epsilon_{ikl}n_k J_{lj}& = \sum_{m=-2}^{2} p_m (\mathbb{T}^{\text{Bu}}_{2m})_{uA}, \\
2 e^{i}_{\langle A}e^{j}_{ B \rangle} \epsilon_{ikl}n_k J_{lj} &= -\frac{1}{2}\sum_{m=-2}^{2} p_m (\mathbb{T}^{\text{B2}}_{2m})_{AB},
\end{align}
where coefficients $p_m$ are identified as,
\begin{align}
p_2 &=(p_{-2})^{*} = \sqrt{\frac{2\pi }{30}} \left(J_{22}-J_{11}+2 i J_{12}\right), \\
p_1 &= -(p_{-1})^{*} = 2\sqrt{\frac{2 \pi }{15}} \left(J_{13}-i J_{23}\right), \\ 
p_0 &=2\sqrt{\frac{\pi }{5}} \left(J_{11}+J_{22}\right).
\end{align}
In terms of the magnetic parity tensor
harmonics, the perturbation is therefore,
\bea
H_{\mu \nu}^{\text{(B)}}\big{|}_{l=2}  &=&  \sum_{m=-2}^{2}\frac{1}{2}\left[-\frac{r^2}{ L^4} \int^{u} p_m
(u')d u'-\frac{1}{ r}\left(1+\frac{r^2}{L^2}\right)p_{m}+\frac{r^2}{ L^2}\dot{p}_m+ r\ddot{p}_{m}\right](\mathbb{T}^{\mathrm{B2}}_{2m})_{\mu \nu} \nonumber \\ 
& & + 
\sum_{m=-2}^{2}
\left[\frac{3}{2r^{2}}\left(1- \frac{2r^2}{3L^2}\right)p_{m}
+\frac{2}{r}\dot{p}_{m}+\ddot{p}_{m}\right] (\mathbb{T}^{\mathrm{Bu}}_{2m})_{\mu \nu}  \label{28III24.01},
\eea
which satisfies linearised Einstein's equations.

To get to the form of CHK solution \eqref{28III24.01} from dS Teukolsky solutions \eqref{magnetic_teukolsky},  we need to do a gauge transformation \eqref{diffeo}. Specifically, via the gauge transformation we need to set the  coefficient of $(\mathbb{T}^{\mathrm{B1}}_{2m})$ tensor harmonics to zero. Via \eqref{transform_magnetic_2}
we have
\be
\frac{1}{r} \left( r \partial_r \xi^{\mathrm{(B)}}_{2m} - 2 \xi^{\mathrm{(B)}}_{2m} \right) +  \frac{4}{r} \ddot{B}_{m} + \frac{12}{r^2} \left(1+ \frac{r^2}{3L^2}\right) \dot{B}_{m}+ \frac{12}{r^3}\left(1 + \frac{r^2}{3L^2}  \right) B_m=0.
\ee
Integration of this equation gives, 
\be
\xi^{\mathrm{(B)}}_{2m}  =2  \ddot{B}_m(u)  + \frac{4}{r} \left( 1 + \frac{r^2}{L^2}\right)  \dot{B}_m(u) + \frac{3}{r^2} \left( 1 + \frac{2r^2}{3L^2}\right)  B_m(u) + r^2 c_m(u),
\ee
where $c_m(u)$ is an integration `constant'. The choice
\be
c_m(u) =\frac{4 B_m}{L^4},
\ee
with the identification 
\be
B_m(u) = \frac{1}{2}\int^u p_m(u') du' ,
\ee
maps de Sitter Teukolsky waves \eqref{magnetic_teukolsky} to CHK solution \eqref{28III24.01}.

\subsection{Electric parity quadrupole truncation}

The electric parity case is similar but more involved. The electric parity CHK quadrupolar solution is,
\begin{align} 
\label{CHK_E_1}
H_{uu}^{\text{(E)}}\big{|}_{l=2}&= (\delta^{ij}-3n^i n^j)\partial_u (\zeta_{ij}+2H^2 Q^{(\rho+p)}_{ij})  
+\frac{2M}{r}+\frac{2N}{r^2}-\frac{(\delta^{ij} - 3n^i n ^j)Q^{(\rho+p)}_{ij}}{r^3},\\
\label{CHK_E_2}
H_{uA}^{\text{(E)}}\big{|}_{l=2} &= e_A^i n^j\left(-(H^{-2}\partial_u^2 \zeta_{ij}+2\partial_u^2 Q^{(\rho+p)}_{ij})+ \frac{4\partial_u Q^{(\rho+p)}_{ij}}{r}+\frac{ 3 Q_{ij}^{(\rho+p)}}{r^2}\right),\\
H_{AB}^{\text{(E)}}\big{|}_{l=2}&= e^i_{\langle A} e^j_{B \rangle}\left( r^2  \left(\partial_u \zeta_{ij}+2H^2 \partial_u Q^{(\rho+p)}_{ij}\right)+ r  \left( 3\zeta_{ij} +2 (\partial_u^2 -H^2)Q^{(\rho+p)}_{ij}\right) + \frac{2}{r}  Q^{(\rho+p)}_{ij}\right),
\label{CHK_E_3}
\end{align}
where
\begin{align}
N &= -(\delta_{ij}-3 n_{i}n_{j}) \partial_{u} Q_{ij}^{(\rho +p)}, \\
M &= -(\delta^{ij} - 3n^i n ^j)(\partial_{u}^{2}Q_{ij}^{(\rho +p)}-H^{2}Q_{ij}^{(\rho +p)}).
\end{align}

In order to write the electric parity CHK perturbation in terms of the electric parity tensor harmonics, let us consider the three combinations of angular coordinates 
$(\delta^{ij}-3n^i n^j), \ e_A^i n^j, \ e^i_{\langle A} e^j_{B \rangle}
$ that appear in \eqref{CHK_E_1}--\eqref{CHK_E_3}. These three combinations can be written in terms of electric parity tensor harmonics. For an arbitrary symmetric cartesian tensor $\chi_{ij}$ we have,
\bea
\label{dS_electric_CHK_to_tensor_harmonics}
(\delta^{ij}-3n^i n^j) \chi_{ij} &=& \sum_{m=-2}^{2} q_m (\mathbb{T}^{\text{uu}}_{2m})_{uu}, \\
e_A^i n^j \chi_{ij}
&=&-\frac{1}{6} \sum_{m=-2}^{2}  q_m (\mathbb{T}^{\text{Eu}}_{2m})_{uA}, \\
e^{i}_{\langle A}e^{j}_{ B \rangle} \chi_{ij} 
&=&-\frac{1}{12} \sum_{m=-2}^{2}  q_m (\mathbb{T}^{\text{E2}}_{2m})_{AB}, 
\eea
where the coefficients $q_m$ are identified as
\begin{align}
q_2&= (q_{-2})^{*} = \sqrt{\frac{6 \pi }{5}} \left(\chi_{22}-\chi_{11}+2 i \chi_{12}\right), \\
q_1&= (q_{-1})^{*} = 2 \sqrt{\frac{6 \pi }{15}} \left(\chi_{13}-i \chi_{23}\right), \\
q_0&=2 \sqrt{\frac{\pi }{5}} \left(\chi_{11}+\chi_{22}-2 \chi_{33}\right).
\end{align}

As the next step, we consider replacing $Q_{ij}^{(\rho + p)}$ in favour of $\zeta_{ij}$ via \eqref{zeta-eq}. Then, the solution
\eqref{CHK_E_1}--\eqref{CHK_E_3} is written only in terms of $\zeta_{ij}(u)$ alone. We identify 
\be
\chi_{ij}=- L^4\zeta_{ij} =- H^{-4}\zeta_{ij}. \label{chi-ij}
\ee 
Then, the CHK form of the perturbation is, 
\begin{align}
H_{\mu \nu}^{\text{(E)}}\big{|}_{l=2} & =\sum_{m=-2}^{2}\left[\frac{3}{2 L^2 r^3}\left(1-\frac{2 r^2}{L^2}\right) q_m+\frac{3}{L^2 r^2}\left(1-\frac{4 r^2}{3 L^2}\right) \dot{q}_m-\frac{1}{2 r^3}\left(1-\frac{8 r^2}{L^2}\right) \ddot{q}_m
\right. \nonumber \\
& \qquad \qquad  \left.
- \frac{1}{r^2}\left(1-\frac{r^2}{L^2}\right) q_m^{(3)}-\frac{q_m^{(4)}}{r}\right](\mathbb{T}^\mathrm{uu}_{2m})_{\mu \nu} \nonumber\\
&
\quad +\sum_{m=-2}^{2}\left[\frac{3 q_{m}}{4 L^2 r^2}+\frac{\dot{q}_m}{L^2 r}-\frac{1}{4 r^2}\left(1+\frac{8 r^2}{3 L^2}\right) \ddot{q}_m -\frac{q_{m}^{(3)}}{3 r}+\frac{1}{6} q_{m}^{(4)}\right](\mathbb{T}^\mathrm{Eu}_{2m} )_{\mu \nu}\nonumber\\
& \quad +\sum_{m=-2}^{2}\left[\frac{q_{m}}{4 L^2 r}+\frac{r^2 \dot{q}_m}{3 L^4}-\frac{1}{12 r}\left(1-\frac{4 r^2}{L^2}\right) \ddot{q}_m-\frac{r^2 q_{m}^{(3)}}{12 L^2}-\frac{1}{12} r q_{m}^{(4)}\right](\mathbb{T}^\mathrm{E2}_{2m} )_{\mu \nu}, \label{CHK_electric_final}
\end{align}
which satisfies linearised Einstein's equations. The use of the appropriate powers of $L$ in equation \eqref{chi-ij} allows us to take the $L\to \infty$ limit straightforwardly in \eqref{CHK_electric_final}.

We can get the solution \eqref{CHK_electric_final} by applying a diffeomorphism \eqref{diffeo} on the electric parity de Sitter Teukolsky wave. It turns out that this calculation is rather straightforward given our presentation above. We note that the CHK electric parity solution \eqref{CHK_electric_final} has the same set of tensor harmonics as the  BBP electric parity solution \eqref{eq:elecsol_BBP}. As a result, we can simply follow the  computation of section \ref{sec:EP_BBP} up to equation \eqref{relation_integral_constants} without any change. 

As in section \ref{sec:EP_BBP}, in the end, we are left with only one function $c_1$ after setting the coefficients of $
\mathbb{T}^\mathrm{Ru}, \ \mathbb{T}^\mathrm{L0}, \ \mathbb{T}^\mathrm{T0}, \ \mathbb{T}^\mathrm{E1}$ tensor harmonics to zero. Choosing $c_1$ as
\be
c_1=\frac{1}{8 L^4}\left(2 L^2 \ddot{q}_m-8 q_m+L^2 \ddot{A}_m-12 A_m\right),
\ee
and identifying the combination $-\left(9 A_m-L^2\ddot{A}_m\right)$ as $6 q_m-2L^2\ddot{q}_m$, i.e., 
\be
6 q_m-2L^2\ddot{q}_m=-\left(9 A_m-L^2\ddot{A}_m\right),
\ee
maps the electric parity de Sitter Teukolsky waves to electric parity CHK waves. 

\section{Energy flux for de Sitter Teukolsky waves}
\label{sec:energy}

In this section, we compute the energy flux across future timelike infinity for 
linearised de Sitter Teukolsky waves. An expression for the energy flux for linearised perturbation   is given as (see eq.~(30) of \cite{Kolanowski:2020wfg} and eq.~(4.10) of \cite{Ashtekar:2015lxa}),
\be \label{16III24.04}
E=\frac{L}{16\pi G}\int_{\mathcal{I}^{+}} d^3x \sqrt{q^{(0)}} {} \mathcal {E}_{a b}^{(0)} (L^{2} \partial_{u}h_{c d}^{(0)}) {q}^{(0)a c}  {q}^{(0)b d},
\ee
where $q^{(0)}_{ab} $ is the metric on $\mathcal{I}^{+}$ induced from the de Sitter background.  In coordinates $\{u,r,\theta,\phi\}$, the background de Sitter spacetime metric  has the form \eqref{dS_metric}.
The induced metric $q^{(0)}_{ab}$ is therefore 
\be
q^{(0)}_{ab} dx^{a} dx^{b}:=\lim_{r\to \infty} (r^{-2}{ds}^{2}) \bigg{|}_{r=\mathrm{const}} = \frac{1}{L^2} du^{2}+d\theta^2 + \sin^2 \theta d\phi^2.
\ee
In equation \eqref{16III24.04}, $h_{ab}^{(0)}$ is
\be
h_{ab}^{(0)}:=\lim_{r \to \infty} 
(r^{-2} h_{ab}) \bigg{|}_{r=\mathrm{const}}, \qquad  \mbox{with} \qquad a, b \in \{u, \theta, \phi \},
\ee
where $h_{\mu \nu}$ is the linearised perturbation.  $\mathcal{E}_{ab}^{(0)}$ is the rescaled electric part of the  \textit{linearised} Weyl tensor at the boundary,
\be 
\mathcal{E}^{(0)}_{ab}:= \lim_{r \to \infty} r C_{a \rho b \sigma} n^{\rho} n^{\sigma},
\ee
where $n^{\mu}$ is\footnote{There is a minor typo in the expression of $n^{\mu}$ in appendix A of  \cite{Bonga:2023eml}. Throughout this section we only work with $\{u, r, \theta, \phi\}$ coordinates. For the convenience of the reader, we simply translate relevant formulae from \cite{ Anninos:2010zf, Ashtekar:2015lxa, Kolanowski:2020wfg, Bonga:2023eml} into the coordinate-based language without introducing the conformally compactified spacetime.}
\be 
n^{\mu} =  \left\{1, -\frac{1}{L^{2}},0,0 \right\}.
\ee

We can easily compute various pieces required to evaluate equation \eqref{16III24.04} for the magnetic and electric parity de Sitter Teukolsky waves.   A calculation shows that for the magnetic parity de Sitter Teukolsky waves, 
\bea \label{16III24.01}
\mathcal{E}_{ab}^{(0)} dx^{a} dx^{b} =\sum_{m=-2}^{2}\left[ \frac{2}{L^{4}}\partial_{u}^{2}\left(4B_{m}-L^{2} \ddot B_{m}\right) X_{A}^{2m} {}du dx^{A}+\frac{1}{L^{2}}\partial_{u}^{3}\left(4B_{m}-L^{2} \ddot B_{m}\right) X_{AB}^{2m} {} dx^{A} dx^{B}\right].
\eea
The non-vanishing $h_{ab}^{(0)}$ components are $h_{uA}^{(0)}$ and $h_{AB}^{(0)}$.
From \eqref{magnetic_teukolsky}, we have
\bea \label{data1}
h_{uA}^{(0)}&:=&\lim_{r\to \infty} (r^{-2} h_{uA})= \frac{4}{L^{4}} \sum_{m=-2}^{2} \dot B_{m} X_{A}^{2m}, \\ \label{data2}
h_{AB}^{(0)}&:=& \lim_{r\to \infty} (r^{-2} h_{AB})= \frac{2}{L^2} \sum_{m=-2}^{2} \bigg(\ddot B_{m}+\frac{3}{L^2} B_{m}\bigg) X_{AB}^{2m}.
\eea
Upon substituting \eqref{16III24.01}--\eqref{data2} in \eqref{16III24.04}, we obtain the energy flux to be,
\bea \label{16III24.05}
E^{\mathrm{(B)}}=\frac{3}{2\pi G} \sum_{m=-2}^{2}\int_{-\infty}^{\infty} du \left[ (B^{(4)}_{m})^{2}+\frac{5}{L^{2}}(B^{(3)}_{m})^{2}+\frac{4}{L^{4}} (\ddot B_{m})^{2}\right].
\eea
In the intermediate steps, we have dropped certain total derivative terms, and have used various properties of the vector and tensor spherical harmonics. We note that our result \eqref{16III24.05} matches with that of BBP \cite{Bonga:2023eml} with the identification $\dot B_{m}=\frac{1}{2}b_{m}$, cf.~\eqref{iden-BBP_mag}. The fact that our energy flux expression matches with that of BBP is no surprise (but still non-trivial). This is  because the Stewart--Walker lemma ensures that the linearised Weyl tensor is gauge-invariant. We already established  in section \ref{sec:BBP} that de Sitter Teukolsky waves are related to BBP waves via a gauge transformation.

A similar but somewhat lengthy calculation gives the energy flux for the electric parity Teukolsky waves,
\bea 
E^{\mathrm{(E)}} &=& \frac{3}{2\pi G} \sum_{m=-2}^{2}\int_{-\infty}^{\infty}du \nonumber \\ 
& &  \left[ \left(\frac{A_{m}^{(5)}}{24}-\frac{3}{8L^{2}}  A_{m}^{(3)}\right)^{2}+\frac{5}{L^{2}}\left(\frac{A_{m}^{(4)}}{24}-\frac{3}{8L^{2}} \ddot A_{m}\right)^{2}+\frac{4}{L^{4}} \left(\frac{A^{(3)}_{m}}{24}-\frac{3}{8L^{2}} \dot A_{m}\right)^{2}\right]. 
\eea
Once again, this result matches with that of \cite{Bonga:2023eml} with the identification $a_{m}=-\frac{3 A_{m}}{4L^{2}} + \frac{\ddot A_{m}}{12}$, cf.~\eqref{iden_a_A}. For both the electric and magnetic parity waves, the energy flux across future timelike infinity $\mathcal{I}^{+}$ is manifestly positive. Note that the energy flux is positive for any infinitesimal portion of future timelike infinity. 
\section{Conclusions}
\label{sec:Conclusions}

In this paper, we presented a detailed study of linearised quadrupole gravitational waves in de Sitter spacetime. There were already several results in the literature, but the interrelations between them were not clear. An important aim of our paper was to bring together various results and establish interrelations between them. In doing so, we found it most coherent to discuss de Sitter Teukolsky waves and relate these solutions to other results. de Sitter Teukolsky waves are linearised quadrupole gravitational waves (solutions to homogeneous linearised Einstein equations) in the transverse-traceless gauge in de Sitter   spacetime. We presented these solutions in section \ref{sec:dS_Teukolsky}.   A key feature of our presentation is that in the cosmological constant going to zero limit, our solutions reduce to Teukolsky solutions in flat spacetime \cite{Teukolsky:1982nz}. In order to find these solutions, we do not need to resort to other formalisms. As in Teukolsky's work \cite{Teukolsky:1982nz}, using the tensor harmonic decomposition and making appropriate ansatz for the functions appearing as coefficients of various tensor harmonics, we  could solve linearised Einstein equations.

In later sections, we exhibited gauge transformations  starting from de Sitter Teukolsky solutions that related our solutions to other de Sitter gravitational wave solutions reported in the literature. Specifically, in section \ref{sec:BBP}, we showed that our linearised solutions are equivalent to the recently reported Bonga, Bunster, and P\'erez \cite{Bonga:2023eml} linearised solutions in Bondi gauge.   In section \ref{sec:ls_construction}, we showed that our linearised solutions are also equivalent to quadrupolar gravitational perturbations constructed by Loganayagam and Sheyte \cite{Loganayagam:2023pfb, Loganayagam:2023pfb_2}.  In section \ref{sec:CHK_dS}, we explored in detail the relation of our solutions  
to the recent paper of  Comp\`ere, Hoque, and Kutluk (CHK) 
\cite{Compere:2023ktn}.  Expanding on the work of Ashtekar, Bonga, and Kesavan (ABK) \cite{deVega:1998ia, Ashtekar:2015lla,  Ashtekar:2015ooa, Ashtekar:2015lxa}, CHK obtained expressions for the metric perturbations around de Sitter generated by localised sources. They also made more precise the meaning of the quadrupolar truncation in the multipolar expansion. In section \ref{sec:CHK_dS}, we showed that our linearised solutions are equivalent to their solutions. The fact that our quadrupolar solutions precisely match with CHK solutions, is a highly non-trivial consistency check not only on our calculations and but also on the arguments made in \cite{Compere:2023ktn} for a consistent quadrupolar truncation. These results are small steps towards a complete theory of  ``post-de Sitter'' multipolar expansion. Finally, in section \ref{sec:energy}, we computed energy flux for de Sitter Teukolsky waves. 

The results presented in the current paper offer several opportunities for future research. An important feature of our quadrupolar solutions in section \ref{sec:dS_Teukolsky} is that each metric function in the $\{u, r, \theta, \phi \}$ coordinates has only a finite number of terms when expanded in powers of $r$. Given our discussion in section \ref{sec:ls_construction}, it is clear that this feature continues to be true at higher multipolar  order.  It will be a useful exercise to explicitly write de Sitter Teukolsky solutions  for general $l$, such that on one hand they have a limit to Rinne solutions~\cite{Rinne:2008ig} in the cosmological constant going to zero limit  and on the other hand they are related to    gravitational perturbations of section \ref{sec:ls_construction}  via a linearised diffeomorphism.

Our results also suggest that as far as linearised solutions are concerned, Bonga-Bunster-P\'erez \cite{Chrusciel:2020rlz, Chrusciel:2020rlz2, Bonga:2023eml} boundary conditions and  Comp\`ere-Fiorucci-Ruzziconi 
\cite{Compere:2019bua, Compere:2020lrt} 
boundary conditions are equally good to work with. We wonder whether a study of non-linear solutions can shine additional light on the issue of which boundary conditions are more physical to work with. In related studies, Fern\'andez-\'Alvarez and Senovilla \cite{Fernandez-Alvarez:2020hsv, Fernandez-Alvarez:2021yog, Fernandez-Alvarez:2021uvz, Senovilla:2022pym} have focused on how to identify the presence of gravitational radiation in  de Sitter spacetime using geometric tools only.  Their criterion is based on the super-Poynting vector at future timelike infinity. We wonder if such ideas can help in clarifying the issue of the boundary conditions. In the same vein, it is timely to explore how the asymptotic symmetries of refs.
\cite{Compere:2019bua, Compere:2020lrt} are to be understood in the framework of  refs.\cite{Chrusciel:2020rlz, Chrusciel:2020rlz2, Bonga:2023eml} and vice-versa.  We hope to contribute to this discussion in our future work.

\subsection*{Acknowledgements}

We thank K G Arun, B\'eatrice Bonga,  Geoffrey Comp\`ere, Bala Iyer, R.~Loganayagam, Marc Geiller, and Omkar Shetye  for discussions and help with various technical issues that made this project possible. 
We also thank Geoffrey Comp\`ere,  Ghanashyam Date, K.~Narayan,  and Omkar Shetye for comments on the manuscript. 
The work of AV is partly supported by SERB Core Research Grant CRG/2023/000545 and by the ``Scholar in Residence'' program of IIT Gandhinagar. The work of JH is supported in part by a MSCA Fellowship 
CZ -- UK2 (reg.~no. CZ.02.01.01/00/$22\_010$/0008115) from the Programme Johannes Amos Comenius co-funded by the European Union. JH also acknowledges  support from the Czech Science 
Foundation Grant 22-14791S.

\appendix

\section{Zerilli tensor harmonics}
\label{sec:Zerilli}
Out of the ten Zerilli tensor harmonics, three are magnetic parity and seven are electric parity. In this paper, the normalisation of these tensor harmonics does not play an important role. 
We use them to capture the angular dependence in a convenient manner. Rinne \cite{Rinne:2008ig} in his generalisation of Teukolsky solutions to higher multipoles uses spin-weighted harmonics, which are closely related to Zerilli tensor harmonics \cite[Section 2]{Thorne:1980ru}. Define,
\bea
X(f) &=& 2 \partial_\theta \partial_\phi f - 2 \cot \theta \partial_\phi f,\\
W(f) &=& \partial^2_\theta f - \cot \theta \partial_\theta f - \frac{1}{\sin^2 \theta } \partial^2_\phi f.
\eea
We use $\{u, r, \theta, \phi \}$ coordinates. Then, the three magnetic parity tensor harmonics are \cite[Chapter 3]{Maggiore:2007ulw} \cite[Chapter 12]{Maggiore:2018sht}:
\be
(\mathbb{T}^\mathrm{Bu}_{lm} )_{\mu \nu} =\left(
\begin{array}{cccc}
0 & 0 & \frac{1}{\sin \theta} \partial_\phi Y_{lm}& - \sin \theta \partial_\theta Y_{lm} \\
0 & 0 & 0 & 0 \\
\frac{1}{\sin \theta} \partial_\phi Y_{lm} & 0 & 0 & 0 \\
- \sin \theta \partial_\theta Y_{lm} & 0 & 0 & 0\\
\end{array}
\right),
\ee
\be
(\mathbb{T}^\mathrm{B1}_{lm} )_{\mu \nu} =\left(
\begin{array}{cccc}
0 & 0 & 0 & 0 \\
0 & 0 & \frac{1}{\sin \theta} \partial_\phi Y_{lm}& - \sin \theta \partial_\theta Y_{lm} \\
0 &  \frac{1}{\sin \theta} \partial_\phi Y_{lm} & 0 & 0 \\
0 & - \sin \theta \partial_\theta Y_{lm} & 0 & 0\\
\end{array}
\right),
\ee

\be
(\mathbb{T}^\mathrm{B2}_{lm} )_{\mu \nu} =\left(
\begin{array}{cccc}
0 & 0 & 0 & 0 \\
0 & 0 & 0 & 0 \\
0 & 0 & -\frac{1}{\sin \theta}  \,X( Y_{lm})& \sin \theta \, W( Y_{lm}) \\
0 & 0 & \sin \theta  \, W( Y_{lm})&\sin \theta  \, X( Y_{lm}) \\
\end{array}
\right).
\ee
The seven electric parity ones are:
\be
(\mathbb{T}^\mathrm{uu}_{lm} )_{\mu \nu} =\left(
\begin{array}{cccc}
1& 0 & 0& 0 \\
0 & 0 & 0 & 0 \\
0 & 0 & 0 & 0 \\
0 & 0 & 0 & 0 \\
\end{array}
\right) Y_{lm},
%
\quad
(\mathbb{T}^\mathrm{Ru}_{lm} )_{\mu \nu} =\left(
\begin{array}{cccc}
0& 1 & 0& 0 \\
1 & 0 & 0 & 0 \\
0 & 0 & 0 & 0 \\
0 & 0 & 0 & 0 \\
\end{array}
\right) Y_{lm},
\ee
\be
(\mathbb{T}^\mathrm{L0}_{lm} )_{\mu \nu} =\left(
\begin{array}{cccc}
0& 0 & 0& 0 \\
0 & 1 & 0 & 0 \\
0 & 0 &0 & 0 \\
0 & 0 & 0 &0 \\
\end{array}
\right) Y_{lm},
%
\quad (\mathbb{T}^\mathrm{T0}_{lm} )_{\mu \nu} =\left(
\begin{array}{cccc}
0& 0 & 0& 0 \\
0 & 0 & 0 & 0 \\
0 & 0 &1 & 0 \\
0 & 0 & 0 &\sin^2 \theta \\
\end{array}
\right) Y_{lm},
\ee

\be
(\mathbb{T}^\mathrm{Eu}_{lm} )_{\mu \nu} =\left(
\begin{array}{cccc}
0 & 0 & \partial_\theta Y_{lm}&  \partial_\phi Y_{lm} \\
0 & 0 & 0 & 0 \\
\partial_\theta Y_{lm} & 0 & 0 & 0\\
\partial_\phi Y_{lm}& 0 & 0 & 0\\
\end{array}
\right),
\quad 
(\mathbb{T}^\mathrm{E1}_{lm} )_{\mu \nu} =\left(
\begin{array}{cccc}
0 & 0 & 0 & 0 \\
0 & 0 & \partial_\theta Y_{lm}&  \partial_\phi Y_{lm} \\
0 & \partial_\theta Y_{lm} & 0 & 0 \\
0 &\partial_\phi Y_{lm}& 0 & 0\\
\end{array}
\right),
\ee

\be
(\mathbb{T}^\mathrm{E2}_{lm} )_{\mu \nu} =\left(
\begin{array}{cccc}
0 & 0 & 0 & 0 \\
0 & 0 & 0 & 0 \\
0 & 0 &W( Y_{lm})& X(Y_{lm}) \,  \\
0 & 0 & X(Y_{lm})& - \sin^2 \theta \, W( Y_{lm}) \\
\end{array}
\right).
\ee
We work with the following trace-free linear combinations,
\bea\label{S0_from_L0_and_T0}
(\mathbb{T}^\mathrm{S0}_{lm} )_{\mu \nu} &=&\left(1-\frac{r^2}{L^2}\right)^{-1} (\mathbb{T}^\mathrm{L0}_{lm} )_{\mu \nu} - \frac{1}{2}r^2 (\mathbb{T}^\mathrm{T0}_{lm} )_{\mu \nu} \\
&=&
\left(
\begin{array}{cccc}
0& 0 & 0& 0 \\
0 & \left(1-\frac{r^2}{L^2}\right)^{-1} & 0 & 0 \\
0 & 0 & -\frac{1}{2} r^2 & 0 \\
0 & 0 & 0 & -\frac{1}{2} r^2 \sin^2 \theta \\
\end{array}
\right) Y_{lm},
\eea
and
\bea
(\mathbb{T}^\mathrm{ur}_{lm} )_{\mu \nu} &=&(\mathbb{T}^\mathrm{Ru}_{lm} )_{\mu \nu} + 2 \left(1-\frac{r^2}{L^2}\right)^{-1}  (\mathbb{T}^\mathrm{L0}_{lm} )_{\mu \nu} \label{ur_from_L0_and_Ru}\\
&=&
\left(
\begin{array}{cccc}
0& 1 & 0& 0 \\
1 &2 \left(1-\frac{r^2}{L^2}\right)^{-1} & 0 & 0 \\
0 & 0 & 0 & 0 \\
0 & 0 & 0 & 0 \\
\end{array}
\right) Y_{lm}.
\eea
The trace is with respect to the background metric $\bar g_{\mu \nu}$ \eqref{dS_metric_matrix}: $h =  \bar g^{\mu \nu} h_{\mu \nu}$.

\section{Some details on solving linearised Einstein equations}
\label{app:solving_E_eqs}

\begin{table}[t!]
\centering
\begin{tabular}{|c|c|c|c|}
\hline
component $\mu \nu$ & coefficient &  function/constant & how it is determined \\
\hline
\hline
$u\phi$ & $B^{(4)}$ & $f_3$ & differential relation $c^{f_3}$ undetermined constant  \\
$\theta\phi$ & $B^{(4)}$ & $g_2$ & differential relation $c^{g_2}$ undetermined constant  \\
$u\phi$ & $B^{(3)}$ & $f_2$ & differential relation $c^{f_2}$ undetermined constant  \\
$\theta\phi$ & $B^{(3)}$ & $g_1$ & differential relation $c^{g_1}$ undetermined constant  \\
$u\phi$ & $B^{(2)}$ & $f_1$ & differential relation $c^{f_1}$ undetermined constant  \\
$\theta\phi$ & $B^{(2)}$ & $g_0$ & differential relation $c^{g_0}$ undetermined constant  \\
$u\phi$ & $B^{(1)}$ & $f_0$ & differential relation $c^{f_0}$ undetermined constant  \\
$\theta\phi$ & $B^{(1)}$ & $c^{f_3}$ \& $c^{f_2}$ & algebraic relations from different powers of $r$ \\
$u\phi$ & $B$ & $c^{f_1}$ \& $c^{f_0}$ & algebraic relations from different  powers of $r$ \\
\hline
\end{tabular}
\caption{\sl The first row says that the coefficient of $B^{(4)}$ in  the $u\phi$ component of the Einstein equation fixes $f_3$ upto to a constant $c^{f_3}$. This algorithm needs to be followed in the sequential order as listed.}
\label{table:1}
\end{table}

In this appendix, we briefly explain 
how we constructed the magnetic parity de Sitter Teukolsky solution. A similar consideration applies to the electric parity solution, which we do not discuss. We begin by writing appropriate ansatzes for functions  $f_\mathrm{Bu}^{2m}(u,r), f_\mathrm{B1}^{2m}(u,r),$ and $f_\mathrm{B2}^{2m}(u,r)$ appearing in equation \eqref{magnetic_teukolsky}. We restrict ourselves to $l=2, m=0$ and drop the $(l,m)$ indices. Similar considerations apply for $l =2, m \neq 0$. We take,
\bea
f_\mathrm{Bu}(u,r) &=& f_3(r) B^{(3)}(u) + f_2(r) B^{(2)}(u)+  f_1(r) B^{(1)}(u) + f_0(r) B(u) ,  \\
f_\mathrm{B1}(u,r)&=& g_3(r) B^{(3)}(u) + g_2(r) B^{(2)}(u)+  g_1(r) B^{(1)}(u) + g_0(r) B(u) , \\ 
f_\mathrm{B2}(u,r)&=& h_3(r) B^{(3)}(u) + h_2(r) B^{(2)}(u)+  h_1(r) B^{(1)}(u) + h_0(r) B(u),
\eea
where $B^{(k)}(u)$ is the $k$-th $u$ derivative of the function $B(u)$.
When we impose the transversality condition ${\bar \nabla}^\mu h_{\mu \nu}  = 0$ (the traceless property is automatic given the ansatz \eqref{magnetic_teukolsky}) we get only one non-zero component of the equation, namely the $\phi$ component. The highest derivative of $B(u)$ that appears there is $B^{(4)}(u)$. Since $B(u)$ is an arbitrary function, that equation can only be satisfied when individual coefficients of $B(u)$ and its derivatives are all zero. In this way, the transversality condition fixes, $g_3, h_3, h_2, h_1, h_0$. These functions are fixed in an algebraic way. $g_3$ is in fact fixed to zero. $ h_3, h_2, h_1, h_0$ are non-trivial.  The list of undetermined functions are $f_0, f_1, f_2, f_3, g_0, g_1, g_2$.

Linearised Einstein equations give three independent relations from $u\phi$, $r\phi$, $\theta\phi$ equations. These equations start at $B^{(4)}(u)$ derivative. We use $B^{(4)}(u)$, $B^{(3)}(u)$, $B^{(2)}(u)$, $B^{(1)}(u)$, $B(u)$ coefficients to fix the undetermined functions. Some details are in Table \ref{table:1}. From this table we see that $c^{g_0}$, $c^{g_1}$, $c^{g_2}$ are the only three undetermined constants. Choosing 
\begin{align}
c^{g_0} &= \frac{1}{L^2},  & c^{g_1} &= 0, & c^{g_2} &= 4,
\end{align}
gives a solution  with the property that in the large $L$ limit it reduces to the magnetic parity Teukolsky wave in flat space.

\section{Flat spacetime Teukolsky waves in Bondi gauge}

\label{app:CC_zero}

It is instructive to discuss in detail how the Teukolsky solutions \cite{Teukolsky:1982nz} in flat spacetime are related to the quadrupolar solutions, say, as presented in refs.~\cite{Maggiore:2007ulw, Blanchet:2020ngx}. In particular, it is useful to have the identifications between the functions appearing in flat spacetime Teukolsky solutions and the electric and magnetic parity quadrupole  moments. These identifications help in developing intuition for the Teukolsky solutions.

For the purpose of this appendix, the most convenient way to obtain the flat spacetime solutions in terms of the source moments  is to  take the $H \rightarrow0$ limit of the quadrupolar Comp\`ere, Hoque, and Kutluk (CHK)  solutions discussed in  section \ref{sec:CHK_dS}.\footnote{The $H \rightarrow0$ limit is also discussed in \cite{Compere:2023ktn}, equations (3.54)--(3.56).} Restricting ourselves to purely quadrupolar truncation, we get,
\begin{align} 
h_{uu} & =\frac{2}{r}(3n^i n^j -\delta^{ij}) \ddot Q^{(\rho + p)}_{ij} -\frac{2}{r^2} (\delta^{ij}-3n^i n^j)\dot Q^{(\rho +p)}_{ij} +\frac{1}{r^3} (3n^{i}n^{j}-\delta^{ij}) Q_{ij}^{(\rho +p)},\\
h_{uA} &=-2 e^{i}_{A} n^{j}(\ddot Q_{ij}^{(\rho + p)}-  \epsilon_{i k l }n_k\ddot J_{jl})+ \frac{4 e_A^i (n_j\epsilon_{ijk} n_l \dot J_{kl}+n_j\dot Q^{(\rho+p)}_{ij})}{r}+\frac{ 3e_A^i n^j (Q_{ij}^{(\rho+p)}+ \epsilon_{ikl}n_k J_{lj})}{r^2}, \\
h_{AB} &= r^{2} \bigg(\frac{2}{r} e^{i}_{\langle A}e^{j}_{ B \rangle} (\ddot Q_{ij}^{(\rho + p)}- \epsilon_{ikl}n_k \ddot J_{lj}) +\frac{2}{r^{3}}e^{i}_{\langle A}e^{j}_{ B \rangle} ( Q_{ij}^{(\rho + p)}+\epsilon_{ikl}n_k J_{jl})\bigg). 
\end{align}
To relate to Teukolsky solutions it is convenient to separate out the magnetic and electric parity pieces. The discussion below parallel some of the discussion from  section \ref{sec:CHK_dS}, but it is significantly simpler. Since it helps in developing intuition, it deserves a separate presentation.

\paragraph{Magnetic parity.} The magnetic parity truncation is 
\bea \label{mag_flat_1}
h_{uu} &=& 0,\\
h_{uA} &=&2e_A^i n^j n_k \epsilon_{i kl}\left(  \ddot{J}_{jl}+  \frac{2}{r}\dot{J}_{jl}+\frac{3}{2r^2}J_{lj}\right), \\
h_{AB} &=& 2 e^i_{\langle A} e^j_{B \rangle}n_k \epsilon_{ikl}\left(-r \ddot{J}_{lj} + \frac{1}{r} J_{lj}\right). \label{mag_flat_3}
\eea
In order to write this solution  in terms of the magnetic parity tensor harmonics, we consider the combinations $e_A^i n^j  \epsilon_{ikl}n_k J_{lj}$ and $e^{i}_{\langle A}e^{j}_{ B \rangle} \epsilon_{ikl}n_k J_{lj}$. A calculation shows that,
\begin{align}
2 e_A^i n^j\epsilon_{ikl}n_k J_{lj}& = \sum_{m=-2}^{2} b_m (\mathbb{T}^{\text{Bu}}_{2m})_{uA}, \\
2 e^{i}_{\langle A}e^{j}_{ B \rangle} \epsilon_{ikl}n_k J_{lj} &= -\frac{1}{2}\sum_{m=-2}^{2} b_m (\mathbb{T}^{\text{B2}}_{2m})_{AB},
\end{align}
where coefficients $b_m$ are identified as,
\begin{align}
b_2 &=(b_{-2})^{*} = \sqrt{\frac{2\pi }{30}} \left(J_{22}-J_{11}+2 i J_{12}\right), \label{b-iden-1} \\
b_1 &= -(b_{-1})^{*} = 2\sqrt{\frac{2 \pi }{15}} \left(J_{13}-i J_{23}\right), \label{b-iden-2} \\ 
b_0 &=2\sqrt{\frac{\pi }{5}} \left(J_{11}+J_{22}\right). \label{b-iden-3} 
\end{align}
It then follows that in terms of the magnetic parity tensor
harmonics,  perturbation \eqref{mag_flat_1}--\eqref{mag_flat_3} takes the form,
\bea
h_{\mu \nu}^{\text{(B)}} &=&  \sum_{m=-2}^{2}\frac{1}{2}\left[r\ddot{b}_{m} -\frac{1}{ r}b_{m}\right](\mathbb{T}^{\mathrm{B2}}_{2m})_{\mu \nu} 
+
\sum_{m=-2}^{2}
\left[\frac{3}{2r^{2}}b_{m}
+\frac{2}{r}\dot{b}_{m}+\ddot{b}_{m}\right] (\mathbb{T}^{\mathrm{Bu}}_{2m})_{\mu \nu}.  \label{mag_flat_final}
\eea
The choice
\be
\xi^{\mathrm{(B)}}_{2m}  = 2 \ddot{B}_m(u)  + \frac{4}{r}\dot{B}_m(u) + \frac{3}{r^2} B_m(u),
\ee
in equations \eqref{mag_diffeo_1}--\eqref{mag_diffeo_3} maps the magnetic parity Teukolsky wave \eqref{mag_T_flat_1}--\eqref{mag_T_flat_3} to the form \eqref{mag_flat_final} via \eqref{diffeo}
provided we make  the identification 
\be
\dot B_m  = \frac{1}{2}b_m.
\ee 
Through equations \eqref{b-iden-1}--\eqref{b-iden-3}, this identification gives us intuition behind the functions $B_m$ appearing in the Teukolsky wave solutions \eqref{mag_T_flat_1}--\eqref{mag_T_flat_3}.  We note that in the $L \to \infty$ limit, the BBP magnetic parity solution \eqref{28III24.01_BBP} also  gives \eqref{mag_flat_final}. 

\paragraph{Electric parity.} 
The electric parity truncation is 
\begin{align} \label{elec-partity-flat-1}
h_{uu} & =\frac{2}{r}(3n^i n^j -\delta^{ij}) \ddot Q^{(\rho + p)}_{ij} -\frac{2}{r^2} (\delta^{ij}-3n^i n^j)\dot Q^{(\rho +p)}_{ij} +\frac{1}{r^3} (3n^{i}n^{j}-\delta^{ij}) Q_{ij}^{(\rho +p)},\\
h_{uA} &=-2 e^{i}_{A} n^{j}\ddot Q_{ij}^{(\rho + p)}
+ \frac{4 e_A^in_j\dot Q^{(\rho+p)}_{ij})}{r}+\frac{ 3e_A^i n^j Q_{ij}^{(\rho+p)}}{r^2} , \\
h_{AB} &=  2 e^{i}_{\langle A}e^{j}_{ B \rangle}  \left( r \ddot Q_{ij}^{(\rho + p)} +\frac{1}{r} Q_{ij}^{(\rho + p)} \right). \label{elec-partity-flat-3}
\end{align}
In order to write the electric parity perturbation \eqref{elec-partity-flat-1}--\eqref{elec-partity-flat-3}  in terms of the electric parity tensor harmonics, let us consider the three combinations of angular coordinates 
\be
(\delta^{ij}-3n^i n^j), \ e_A^i n^j, \ e^i_{\langle A} e^j_{B \rangle}.
\ee 
For an arbitrary cartesian tensor $\chi_{ij}$ we have,
\bea
(\delta^{ij}-3n^i n^j) \chi_{ij} &=& \sum_{m=-2}^{2} a_m (\mathbb{T}^{\text{uu}}_{2m})_{uu}, \\
e_A^i n^j \chi_{ij}
&=&-\frac{1}{6} \sum_{m=-2}^{2}  a_m (\mathbb{T}^{\text{Eu}}_{2m})_{uA}, \\
e^{i}_{\langle A}e^{j}_{ B \rangle} \chi_{ij} 
&=&-\frac{1}{12} \sum_{m=-2}^{2}  a_m (\mathbb{T}^{\text{E2}}_{2m})_{AB}, 
\eea
where the coefficients $a_m$ are identified as
\begin{align}
a_2&= (a_{-2})^{*} = \sqrt{\frac{6 \pi }{5}} \left(\chi_{22}-\chi_{11}+2 i \chi_{12}\right), \\
a_1&= (a_{-1})^{*} = 2 \sqrt{\frac{6 \pi }{15}} \left(\chi_{13}-i \chi_{23}\right), \\
a_0&=2 \sqrt{\frac{\pi }{5}} \left(\chi_{11}+\chi_{22}-2 \chi_{33}\right).
\end{align}
It then follows that in terms of the electric parity tensor
harmonics,  perturbation \eqref{elec-partity-flat-1}--\eqref{elec-partity-flat-3} takes the form,
\begin{align}
h_{\mu\nu}^{\text{(E)}} & =\sum_{m=-2}^{2}\left[\frac{3}{ r^3} a_m
+ \frac{6}{r^2} \dot{a}_m+\frac{6}{r}  \ddot{a}_m\right](\mathbb{T}^\mathrm{uu}_{2m})_{\mu \nu} 
+\sum_{m=-2}^{2}\left[\frac{3}{2 r^2} a_m +\frac{2}{ r}  \dot{a}_{m} - \ddot{a}_{m}\right](\mathbb{T}^\mathrm{Eu}_{2m} )_{\mu \nu}\nonumber\\
& \quad +\sum_{m=-2}^{2}\left[\frac{1}{ r}a_m+ r  \ddot{a}_{m}\right](\mathbb{T}^\mathrm{E2}_{2m} )_{\mu \nu}, \label{elec_flat_final}
\end{align}
where $\chi_{ij}$ is $-\frac{1}{6}Q_{ij}^{(\rho + p)}$.
The choice 
\bea
\xi^{(u)}_{2m}&=&-\frac{\dot{A}_m}{8 r^3}-\frac{\ddot{A}_m}{4 r^2}-\frac{A_m^{(3)}}{4 r}, \\
\xi^{\mathrm{(R)}}_{2m}&=&-\frac{3 A_m}{8 r^4}-\frac{\dot{A}_m}{2 r^3} -\frac{\ddot{A}_m}{4 r^2}, \\
\xi^{\mathrm{(E)}}_{2m}&=&\frac{A_m}{8 r^3}+\frac{\dot{A}_m}{8 r^2}  +\frac{\ddot{A}_m}{12 r}+\frac{A_m^{(3)}}{12}  r,
\eea
in equations \eqref{elec_diffeo_1}--\eqref{elec_diffeo_3} maps the electric parity Teukolsky waves \eqref{elec_T_flat_1}--\eqref{elec_T_flat_3} to the form \eqref{elec_flat_final} via \eqref{diffeo}
provided we make  the identification 
\be
\ddot{A}_m  = 12 a_m.
\ee 
This identification is the same as in \cite{Teukolsky:1982nz}; it tells us that $\ddot{A}_m$ is directly related to the quadrupole moments $Q_{ij}^{(\rho + p)}$.  We note that in the $L \to \infty$ limit, the BBP electric parity solution \eqref{eq:elecsol_BBP} also  gives \eqref{elec_flat_final}.

\end{document}